\documentclass[twocolumn,trackchanges]{aastex631}

\usepackage{graphicx}
\usepackage{amsmath}
\usepackage{amssymb}
\usepackage{newtxtext,newtxmath}
\usepackage{hyperref}
\usepackage{gensymb}
\usepackage{enumitem}

\newcommand{\Msun}{\ensuremath{M_{\odot}}}
\newcommand{\Gyr}{\ensuremath{\textrm{Gyr}}}
\newcommand{\Myr}{\ensuremath{\textrm{Myr}}}

\newcommand{\kpc}{\ensuremath{\textrm{kpc}}}
\newcommand{\pc}{\ensuremath{\textrm{pc}}}
\newcommand{\kms}{\ensuremath{\textrm{km}/\textrm{s}}}

\newcommand{\FeH}{\ensuremath{[\textrm{Fe}/\textrm{H}]}}
\newcommand{\MgFe}{\ensuremath{[\textrm{Mg}/\textrm{Fe}]}}
\newcommand{\alphaFe}{\ensuremath{[\alpha/\textrm{Fe}]}}

\newcommand{\dex}{\ensuremath{\textrm{dex}}}

\shorttitle{The Milky Way's Phoenix Phase}
\shortauthors{Beane}

\graphicspath{{./}{fig/}}

\begin{document}

\title{Rising from the Ashes: A Metallicity-Dependent Star Formation Gap Splits the Milky Way's $\alpha$-Sequences}

\author[0000-0002-8658-1453]{Angus Beane}
\affiliation{Center for Astrophysics $|$ Harvard \& Smithsonian, Cambridge, MA, USA}

\begin{abstract}
    The elemental abundance distribution of stars encodes the history of the gas-phase abundance in the Milky Way. Without a large, unbiased sample of highly precise stellar ages, the exact timing and nature of this history must be \textit{inferred} from the abundances. In the two-dimensional plane of \alphaFe{}-\FeH{}, it is now clear that two separate populations exist -- the low-$\alpha$ and high-$\alpha$ sequences. We propose that a brief ($\sim300\,\Myr$) halt in star formation within a narrow metallicity bin can lead to a bimodal \alphaFe{} distribution at that metallicity, assuming a rapidly declining gas phase \alphaFe{}. Using simulations of an idealized setup of a high-$z$ galaxy merger, we show that the merger with the Gaia-Sausage-Enceladus satellite at $z\sim2$ is one possible way to trigger such a gap in the Milky Way. This mechanism may also operate in non-merger scenarios. We predict a $\sim300\,\Myr$ gap in stellar ages at a fixed \FeH{} where the $\alpha$-bimodality is prominent ($\FeH\lesssim-0.2$).
\end{abstract}
  

\section{Introduction} \label{sec:intro}
Many elements heavier than hydrogen are produced through nuclear fusion in compact objects such as supernovae, dying low mass stars, and neutron star-neutron star mergers \citep[e.g.][]{2023A&ARv..31....1A}. By necessity, stars inherit the constitutive properties of the gas from which they formed. Moreover, the surface abundance of most elements for most stars do not change over most of their lifetime. By analyzing the surface abundances of stars, we can reconstruct the historical gas-phase composition of a galaxy.

The enrichment of the gas-phase of a galaxy is determined by a complicated combination of physical processes - stellar evolution and supernovae, gas accretion, mergers, gas outflows from stellar and active galactic nuclei (AGN) feedback, metal mixing and diffusion, etc. Because the processes which give rise to this distribution are complex, there is almost certainly some structure in the stellar abundance distribution for every galaxy. However, it has only been definitively measured in the Milky Way, with conflicting claims of detection \citep{2023ApJ...956L..14K} and non-detection \citep{2024IAUS..377..115N} in M31.

The distribution of elemental abundances is a high dimensional space \citep[e.g., 32 elements in][]{2024ApJ...961L..41J}. However, this space is highly degenerate, and so the effective number of dimensions is much smaller -- even possibly compressed to just \FeH{} and age \citep{2019ApJ...883..177N}. Two elements have received particular interest - Fe and elements produced by the $\alpha$-process. Type Ia and Type II supernovae are the main contributors of Fe and $\alpha$ enrichment. Fe is broadly produced in both types, and so its abundance is a proxy for the total metallicity of a star. On the other hand, $\alpha$-elements are mainly produced in Type II supernovae. The ratio of $\alpha$-elements to Fe (\alphaFe{}) is then a measure of the relative contributions of Type Ia and II SNe to the enrichment of a parcel of gas, which typically declines with time \citep{1979ApJ...229.1046T,1986A&A...154..279M}. It has therefore become common to compress the high-dimensional abundance space to the two dimensional \alphaFe{}-\FeH{} plane.

These abundances can be used to decompose the Milky Way's disk, which has a long history dating back to the work of \citet{1983MNRAS.202.1025G}, who noted that the vertical distribution of stellar altitudes is well-fit by a double exponential. This led naturally to a ``thin'' and ``thick'' disk, whose membership can be reasonably determined through kinematics \citep[e.g.][]{2003A&A...410..527B}. It was quickly realized that the thick disk is more $\alpha$-enhanced than the thin disk \citep{1996ASPC...92..307G,1998A&A...338..161F}.

Later studies showed that the disk could be decomposed into high- and low-$\alpha$ sequences without kinematic selection \citep{2011A&A...535L..11A,2012A&A...545A..32A}.\footnote{\citet{2003A&A...410..527B} briefly noted that the thin and thick disk seemed to not overlap in chemistry.} The high-$\alpha$ sequence is older, more centrally compact, and more vertically extended than the low-$\alpha$ sequence \citep{2013A&A...560A.109H,2024IAUS..377..115N}. Although the thick disk is more $\alpha$-enhanced than the thin disk, it is not immediately obvious that the chemical and kinematic separations arise from the same physical process \citep[or that they even exist, see][]{2012ApJ...751..131B}.

Naturally, many different processes that could lead to structure in the abundance plane have been discussed in the literature. An early explanation of the bimodality is based on the two-phase gas infall model \citep{1997ApJ...477..765C,2009IAUS..254..191C,2017MNRAS.472.3637G,2019A&A...623A..60S}. In this model, the thick disk first forms rapidly from an initial infall of gas. Because the typical SFR is high, these stars are $\alpha$-enhanced. In some variants, star formation halts completely before a second supply of pristine gas falls into the Galaxy \citep[][and references therein]{2024arXiv240511025S}. This dilutes the gas supply from which the thin disk forms more gradually, creating a loop feature in the abundance plane. The thin disk is then more $\alpha$-poor because its associated SFR is lower, and in certain scenarios two chemically distinct disks are formed.

A later argument by \citet{2021MNRAS.501.5176K} asserts that the two sequences follow from two phases of gas infall, except driven by stellar feedback instead of cosmological inflow. An initial bursty phase follows from the direct collapse of the gaseous halo. The disk has a high SFR leading to the formation of the high-$\alpha$ sequence. Feedback then halts the inflow, and a slower accretion of high-angular momentum and metal-rich gas commences, forming the low-$\alpha$ sequence.

Another mechanism to generate structure in the abundance plane was pointed out by \citet{2009MNRAS.396..203S}, further developed by \citet{2021MNRAS.507.5882S,2023MNRAS.523.3791C}, and explored by \citet{2011ApJ...737....8L,2021MNRAS.508.4484J}. This model claims that, since stars are thought to migrate from their birth radius, there will be stars throughout the entire disk that formed in the inner disk. These $\alpha$-enhanced stars will then form the high-$\alpha$ sequence. This model and its variants also match some chemodynamic properties of the disk. One salient feature of these models is that the bimodality can result from a smooth star formation history.

Yet another explanation, which also invokes an internal process, is that the formation of clumps at high redshift are responsible for both the chemistry and dynamics of the high-$\alpha$ sequence \citep{2019MNRAS.484.3476C,2020MNRAS.492.4716B,2021MNRAS.502..260B,2023ApJ...953..128G}. Instabilities are thought to form clumps in gas-rich disks, and such clumps are seen at intermediate redshifts \citep[$z\sim2$;][]{2005ApJ...627..632E,2007ApJ...658..763E}. These clumps then self-enrich, forming $\alpha$-enhanced stars. The high-$\alpha$ sequence stops forming once the gas fraction is low enough for the instabilities to no longer arise. This model predicts that the high-$\alpha$ and low-$\alpha$ sequences form simultaneously.

Next, we turn to models which argue the bimodality results from some external influence. Early arguments were made that both the $\alpha$-enhancement of the disk and the thickening of the disk can result from gas-rich mergers \citep{2004ApJ...612..894B,2005ApJ...630..298B,2007ApJ...658...60B,2010MNRAS.402.1489R}.\footnote{See also \citet{2009MNRAS.400.1347C} for an argument invoking semi-analytic models.} These mergers lead to an enhanced SFR which leads to the $\alpha$-enhancement of the thick disk, with \citet{2015A&A...578A..87S} being the first to attempt to explain abundance substructure with a merger.

In cosmological simulations, which naturally include early gas-rich mergers, the situation is not as clear. Early work by \citet{2012MNRAS.426..690B} found a general separation between the thin and thick disk, though other authors found a smooth evolution \citep{2013A&A...558A...9M}. \citet{2018MNRAS.474.3629G} found what they referred to as a chemical dichotomy, and argued that it can come from either gas-rich mergers as described before or a ``compaction'' of the disk (we will return to this point in Section~\ref{ssec:cosmo}). Other authors highlight the metal content of the infalling gas, stating that the metal-poor gas associated with satellites can suddenly dilute or reset the disk's metallicity \citep{2020MNRAS.491.5435B,2024MNRAS.528L.122C}. This interpretation can also be understood in the framework of the two-infall models.

The merger explanation of the bimodality is highly synergistic with our picture of the hierarchical assembly of the stellar halo \citep{2005ApJ...635..931B}. Indeed, there is strong evidence that the Milky Way underwent a significant merger with the so-called Gaia-Sausage-Enceladus satellite \citep[GSE;][]{2018MNRAS.478..611B,2018Natur.563...85H,2020ApJ...901...48N}. This merger is thought to have occurred $\sim8-10\,\Gyr$ ago \citep[see also][]{2020ApJ...897L..18B}. A merger origin of the abundance bimodality is also attractive because it can simultaneously explain the origin of the kinematic thin and thick disk \citep{1985AJ.....90.2015G,1986ApJ...309..472Q,1993ApJ...403...74Q}.

Claims in the literature on the stellar mass of GSE vary widely. Early estimates argued from $6\times10^8$ up to even $10^{10}\,\Msun$ \citep{2018MNRAS.478..611B,2018Natur.563...85H,2019MNRAS.484.4471F,2019MNRAS.487L..47V,2019MNRAS.488.1235M,2020MNRAS.493.5195D,2020MNRAS.497..109F}. Later estimates have been more conservative ranging from a mass of $2.7\times10^8\,\Msun$ to $10^9\,\Msun$ \citep{2019MNRAS.482.3426M,2020MNRAS.492.3631M,2020MNRAS.498.2472K,2021ApJ...923...92N,2022AJ....164..249H}, and even as low as $1.5\times10^8\,\Msun$ \citep{2023MNRAS.526.1209L}.

In this work, we propose that a brief $\sim300\,\Myr$ interruption in the formation of stars at a given metallicity can lead to the formation of an $\alpha$-abundance bimodality at that metallicity. It is not common to study the star formation rate at a specific metallicity, but dividing the stellar population into narrow abundance populations can be a powerful tool \citep[e.g.][]{2012ApJ...753..148B,2012ApJ...751..131B}.

Our proposal assumes that the gas phase's \alphaFe{} is declining sufficiently rapidly at the time of the interruption. Using a set of idealized simulations which mimic the $z\sim2$ merger between the Milky Way and GSE, we show that such a merger can drive the formation of this gap and thus the bimodality. While we demonstrate this mechanism in the context of a merger scenario, it is important to note that our proposal does not inherently require a merger to induce this metallicity-dependent star formation gap. This scenario predicts a $\sim300\,\Myr$ gap in stellar ages at metallicities where the bimodality exists ($\FeH\lesssim-0.2$).

In Section~\ref{sec:methods}, we describe our setup. In Section~\ref{sec:results}, we present in detail the main results of two example simulations before expanding our results to the full suite. In Section~\ref{sec:discussion}, we discuss and interpret our results, as well as connections to previous and future work, before concluding in Section~\ref{sec:conclusion}. Throughout this work we refer to the standard native time unit $\kpc/\left(\kms\right)$ as \Gyr{} for convenience.

\section{Methods}\label{sec:methods}
\subsection{Isolated Setup}\label{ssec:iso_setup}
\begin{figure}
    \centering
    \includegraphics[width=\columnwidth]{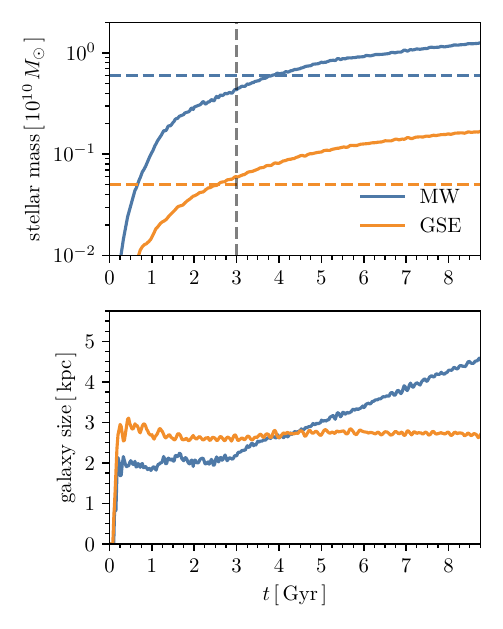}
    \caption{The mass and size evolution of the central (Milky Way, blue) and satellite (GSE, orange) galaxies simulated in isolation. The mass is taken to be the stellar mass within twice the half-mass radius, and the size is taken to be the half-mass radius. In the upper panel, we also show as a horizontal line the mass of the Milky Way's disk and GSE from the best-fit model of \citet{2021ApJ...923...92N}. This comparison is taken to be made at $3\,\Gyr$ (vertical dashed line), our proxy for $z\sim2$. A precise match is not attempted given the wide ranging uncertainties.}
    \label{fig:mass_size}
\end{figure}

We use a modified version of the \texttt{MakeNewDisk} variant described in \citet{2023MNRAS.tmp.2070B}. In isolation, each of the central and satellite galaxies are a compound halo setup, with a \citet{1990ApJ...356..359H} dark matter halo and a gaseous halo with a $\beta$-profile:
\begin{equation*}
\rho = \rho_0 \left[1 + \left(\frac{r}{r_c}\right)^2\right]^{-\frac{3\beta}{2}}
\end{equation*}
The total mass within the virial radius is kept fixed, and the mass of the dark matter halo and central density of the gaseous halo are chosen to satisfy a given baryon fraction $f_b$ within the virial radius. The dark matter halo is initialized to be in gravitational equilibrium with the total potential. The gaseous halo is in gravito-hydrostatic equilibrium, where the temperature is allowed to vary as a function of radius. The azimuthal velocity of the gaseous halo is given as a fraction of the circular velocity. There is no initial stellar disk or bulge, and the gas is initially metal-free. Thus, all star particles and metals are formed self-consistently.

\begin{figure}
    \centering
    \includegraphics[width=\columnwidth]{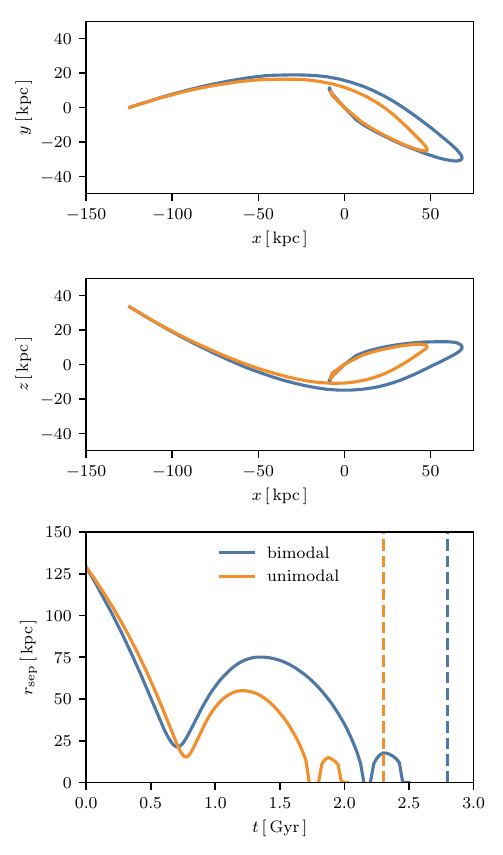}
    \caption{The orbits of the bimodal (blue) and unimodal (orange) simulations. The upper and middle panels show the orbit in the $x$-$y$ and $x$-$z$ planes, respectively. The bottom panel shows the separation distance as a function of time. The orbit begins retrograde but then radializes after the first pericentric passage. The satellite then coalesces quickly after the second pericentric passage, after $\sim2\,\Gyr$ of evolution. A blue dashed line is shown for the bimodal simulation at $2.8\,\Gyr$, the time of a star formation gap at $\FeH=0$ (see Figure~\ref{fig:before_after}). An orange dashed line is shown for the unimodal simulation (which has no gap) at the same time after its second pericentric passage ($2.3\,\Gyr$).}
    \label{fig:orbit}
\end{figure}

We used the fiducial halos in \citet{2021ApJ...923...92N} as a starting point for each galaxy. We then manually varied the different model parameters until we arrived at a setup that resulted in reasonable galaxies as determined by their stellar mass. For the central (Milky Way) galaxy, we set $M_{200}=5\times10^{11}\,\Msun$, $c_{200}=4.1$, $\beta=0.8$, $r_c=9\,\kpc$, $f_b=0.08$, and $v_{\phi}/v_{\textrm{c}}=0.2$, where $c_{200}$ is the concentration and $v_{\phi}/v_{\textrm{c}}$ is the azimuthal velocity of the gaseous halo as a fraction of the local circular velocity. For the satellite (GSE) galaxy, we set $M_{200}=2.2\times10^{11}\,\Msun$, $c_{200}=4.33$, $\beta=0.8$, $r_c=6.5\,\kpc$, $f_b=0.06$, and $v_{\phi}/v_{\textrm{c}}=0.4$.

We used a mass resolution of $6\times10^4\,\Msun$ for the gas and $3\times10^5\,\Msun$ for the dark matter. This is closest to a level~4 resolution in the AURIGA simulations \citep{2017MNRAS.467..179G}, and is about $0.7\times$ the mass resolution of TNG50-1 \citep{2019MNRAS.490.3234N,2019MNRAS.490.3196P}. All collisionless particles have a fixed softening length of $40\,\pc$. The gas has a softening length $2.5\times$ the cell size, with a minimum size of $10\,\pc$. Snapshots were saved at intervals of $25\,\Myr$.

The stellar mass build-up of our Milky Way-like and GSE-like galaxies is given in Figure~\ref{fig:mass_size}. The upper panel shows the stellar mass history. We attempt to match the expected mass of the present-day thick disk \citep[$\sim6\times10^9\,\Msun$, horizontal blue dashed line][]{2016ARA&A..54..529B} at an evolution time of $\sim3\,\Gyr$ (corresponding to $z\sim2$, vertical dashed gray line).\footnote{Of course, this neglects the significant mass contribution of the bulge, which presumably formed earlier. However, our setup does not form a strong spheroidal component. Using the trick in e.g. \citet{2022MNRAS.515.1524Z}, we take the bulge mass to be twice the counter-rotating stellar mass. At $3\,\Gyr$ in the isolated Milky Way-like galaxy, the bulge mass is $\sim7\times10^{8}\,\Msun$, or $\sim13\%$ of the total mass. The Milky Way's bulge is  $\sim1.5\times10^{10}\,\Msun$, although there is strong debate about just how much of the bulge is a classical bulge which formed before the disk \citep{2016ARA&A..54..529B}. In any case, we did not attempt to match any particular property of the bulge, though one could promote bulge formation by reducing the rotation of the gas in the inner region.} We get reasonably close at $\sim5\times10^9\,\Msun$ (blue line). For GSE, we use the best-fit mass from the $N$-body simulations of \citet{2021ApJ...923...92N} -- $5\times10^8\,\Msun$ (horizontal dashed orange line). For this, we slightly overestimate at $\sim6\times10^8\,\Msun$ (orange line).

As for the galaxy sizes, there is significant spread amongst the real galaxy population, and the sizes are thought to be influenced by the merger history not present in our setup \citep[e.g.][]{2014ApJ...788...28V}. We note that the sizes of each simulated galaxy (lower panel) are within the range of observed galaxy sizes. For the Milky Way, we know the thick disk has scale length of $\sim2\,\kpc$, which converts to a half-mass radius of $\sim3.36\,\kpc$. At $\sim3\,\Gyr$, our Milky Way-like galaxy has a half-mass radius of $\sim2\,\kpc$. Curiously, after $3\,\Gyr$, the size of the Milky Way-like galaxy continues to grow while the GSE galaxy's size remains constant for the duration of the simulation.

\subsection{Orbital Configuration}\label{ssec:orbit_setup}
In order to combine the galaxies, we follow \citet{2021ApJ...923...92N}, and place the satellite on a retrograde orbit. In the fiducial simulation of \citet{2021ApJ...923...92N}, the satellite is placed at the virial radius ($R_0=129\,\kpc$), with the virial velocity ($V_0=129\,\kms$), and with a circularity of $\eta=0.5$. To test minor changes to the orbit, we ran a grid of simulations with $\pm10\%$ in each the starting radius and velocity, and $\pm0.1$ in the circularity, for a total of $27$ simulations. We performed each simulation for a duration of $8\,\Gyr$, and used \texttt{FOF} and \texttt{SUBFIND} in order to identify substructure \citep{2005Natur.435..629S,2009MNRAS.399..497D}.

Some of the simulations in this orbital grid resulted in bimodal abundance distributions, while some had little to no structure in the abundance distribution plane. We will first study two representative simulations in detail chosen based on their structure in the abundance plane as shown in Figure~\ref{fig:fig1}, one which we refer to as bimodal and one as unimodal. For the bimodal simulation, we chose the simulation with $R_0=129\,\kpc$, $V_0=142\,\kms$, and $\eta=0.4$. For the unimodal simulation, the parameters are the same except that $V_0=116\,\kms$. These simulations will later be identified as having the highest and second lowest bimodality score $\mathcal{B}$. We will then examine the full simulation suite, and show the detailed abundance plane for the full suite in Appendix~\ref{app:allmerge}.

We show the bimodal and unimodal simulations' orbits in Figure~\ref{fig:orbit}. We use a shrinking spheres center of mass method to identify the centers of the central and satellite galaxy \citep[e.g.,][]{2003MNRAS.338...14P}.\footnote{The position of the minimum potential particle in each substructure identified by \texttt{SUBFIND} is used as the starting guess, and we use an initial/final radius and step factor of $10\,\kpc$, $5\,\kpc$, and $0.9$, respectively.} The upper and middle panels show the orbits in the $x$-$y$ and $x$-$z$ planes, respectively. The lower panel shows the separation distance as a function of time. The orbit is initially retrograde, but quickly radializes after the first pericentric passage. Coalescence occurs rapidly after the second pericentric passage at $\sim2\,\Gyr$, and \texttt{SUBFIND} ceases to recognize the satellite as a separate subhalo.

\subsection{Feedback and Enrichment Model}\label{ssec:gfm}
Our feedback model is a variant of the Illustris TNG model \citep{2013MNRAS.436.3031V,2017MNRAS.465.3291W,2018MNRAS.473.4077P}. In this model, gravity and magnetohydrodynamics are solved using a \citet{1986Natur.324..446B} tree coupled to a second order finite volume fluid solver in AREPO \citep{2010MNRAS.401..791S,2016MNRAS.455.1134P}. Stellar feedback is included through a subgrid wind particle model \citep{2003MNRAS.339..289S}. AGN feedback follows a dual kinetic and thermal mode for low- and high-accretion rates \citep{2017MNRAS.465.3291W}, though in our setup the AGN is only ever in the high-accretion mode. The central galaxy is seeded with a black hole with the typical seed mass ($8\times10^5\,\Msun$). 

In this work, we made some simplifications to this model in order to aid interpretation. First, we ignore magnetic fields. This was motivated by an initial desire to understand the CGM accretion rates in terms of idealized cooling flow solutions, but we did not revisit turning them back on. In any case, it is not clear if the magnetic fields would be realistically generated given our initial setup. Second, we use a gentler wind feedback model as described in \citet{2019MNRAS.489.4233M}. Because our setup includes both an initially steep central potential and no steady-state disk, a stronger feedback model would require a higher central gas density to achieve a reasonable SFH which introduced its own set of pathological instabilities.

In this model, star particles synthesize elements through three different channels for which we cite the relevant yield tables: SNe Ia \citep{1997NuPhA.621..467N}, SNe II \citep{1998A&A...334..505P,2006ApJ...653.1145K}, and AGB stars \citep{2010MNRAS.403.1413K,2014MNRAS.437..195D,2014ApJ...797...44F}. Each star particle, which is modeled as a simple stellar population, continuously injects metals into its surroundings in the following sequence\footnote{The kinetic/thermal feedback component is handled through the wind generation, which is completely separate in this model.}:
\begin{enumerate}
    \item $t\lesssim10\,\Myr$: no metal injection as the first supernova ($M\sim100\,\Msun$) has not gone off
    \item $10\,\Myr \lesssim t \lesssim 40\,\Myr$: metal injection as $8\,\Msun<M<100\,\Msun$ stars die as Type II SNe
    \item $t\gtrsim40\,\Myr$: metal injection from Type Ia SNe and AGB stars
\end{enumerate}
There are a few things to note about this model: (1) The exact timings are metallicity-dependent. (2) The \MgFe{} of ejected gas from Type II SNe is mass/time-dependent, with more massive stars contributing more Mg than less massive stars. (3) In a Hubble time, type II SNe contribute the vast majority of Mg ($\sim10\times$ AGB and $\sim100\times$ Type Ia SNe). Type Ia and Type II SNe contribute approximately equal amounts of Fe (each $\sim3\times$ AGB). See Figure~1 from \citet{2018MNRAS.473.4077P}. (4) The number of Type Ia SNe is greater for a younger stellar population, with a power law relationship $\propto \left(t/\tau_8\right)^{-1.12}$, where $\tau_8=40\,\Myr$ is the lifetime of an $8\,\Msun$ star.

\subsection{Observed Abundances}\label{ssec:obs_abund}
Our aim in this work is to demonstrate the feasibility of a mechanism for structure formation in the abundance plane. We are only making a qualitative comparison to data. Therefore, we use the ASPCAP DR17 catalog of stellar abundances \citep[][J.A.~Holtzman et al., in preparation]{2016AJ....151..144G}, which is publicly available, well-established, and widely used.

We applied quality cuts and restricted our sample to giants, requiring:
\begin{itemize}[noitemsep]
    \item $\textrm{SNR} > 200$,
    \item $\textrm{VSCATTER} < 1\,\kms$,
    \item STARFLAG not set,
    \item $\varpi/\sigma_{\varpi} > 1$,
    \item $\log{g} < 3.5$,
    \item $\sigma_{\log{g}} < 0.2$,
\end{itemize}
where $\varpi$ is the parallax. We use the parallax, proper motion, and radial velocity from Gaia EDR3 \citep{2016AA...595A...1G,2021AA...649A...1G,2021AA...649A...2L,2021AA...653A.160S}.

We next make a solar neighborhood selection of stars based on their angular momenta. We assume the solar radius and azimuthal velocity are $R_0=8\,\kpc$ and $V_0=220\,\kms$ \citep{2016ARA&A..54..529B}, and select stars which have $L_z$ within $10\%$ of the solar angular momentum. We further require that $\left|z\right| < 3\,\kpc$. As is typically done, we use \FeH{} as an indicator of the total metallicity of a star. We use Mg alone as a representative of the $\alpha$-elements.

For stellar ages, we used the APOKASC-3 catalog \citep{2025ApJS..276...69P}. This catalog uses a combination of APOGEE spectroscopic parameters and \textit{Kepler} time series photometry to compute astroseismic ages. Using only stars with $25\%$ age uncertainties (taken as the maximum of the upper and lower uncertainty), we cross-match this catalog to our larger sample from ASPCAP which results in a sample of 2525 stars.

\subsection{Solar Neighborhood in Simulations}\label{ssec:solarneigh}
When comparing galaxy simulations to the observed solar neighborhood, some ambiguity arises in how to make a ``solar neighborhood-like'' selection of star particles. Naturally, this selection is dependent on the posed question, which in this work is the formation of the abundance bimodality. The Sun is known to sit near the end of the thick disk, where the thick and thin disk have comparable surface densities \citep[the ratio of thick-to-thin is $\sim12\%$][]{2016ARA&A..54..529B}. As a result, the abundance bimodality appears most strongly near the Sun -- further inwards the high-$\alpha$ sequence is more dominant and further outwards the high-$\alpha$ sequence vanishes \citep[e.g.,][]{2015ApJ...808..132H}.

We mimic our selection of the solar neighborhood by also making a cut in angular momentum. However, in the simulation, the high-$\alpha$ disk is more compact than in the Galaxy. Therefore, in order to strike a balance between the low-$\alpha$ and high-$\alpha$ disks, we used an angular momentum cut which is $20\%$ that of our assumed solar angular momentum. In particular, we select all star particles with angular momenta within $30\%$ of $0.2\times8\,\kpc\times220\,\kms$ -- as well as requiring $\left| z \right| < 3\,\kpc$. This corresponds to roughly selecting star particles with radii between $2$ and $5\,\kpc$.

\subsection{Bimodality Score}\label{ssec:bim_score}
Given the modest size of our suite, some method for scoring the degree of bimodality for a given 1D distribution is desirable. Tests of whether a distribution is bimodal or unimodal exist -- e.g., the Hartigan dip test \citep{10.1214/aos/1176346577}, but they lack the ability to rank order based on ``bimodalitiness.'' In order to do this, we fit a given distribution of \alphaFe{} as a two-component Gaussian mixture model. The bimodality score $\mathcal{B}$ is then computed as
\begin{equation}\label{eq:bimscore}
\mathcal{B} = \frac{\left|\mu_1-\mu_2\right|}{\sqrt{\sigma_1^2-\sigma_2^2}}\times w(A_2, t, k)\textrm{,}
\end{equation}
where $(\mu_1, \sigma_1)$ is the mean and standard deviation of the higher weighted mode, $(\mu_2, \sigma_2)$ likewise for the less weighted mode, $A_2$ is the amplitude of the less weighted mode, and $w$ is a penalty function with parameters $t$ and $k$ given by
\begin{equation}\label{eq:bimscore_penalty}
w(A, t, k) = \left[1+e^{-k\left(A-t\right)}\right]^{-1}\text{.}
\end{equation}
We take $t$ and $k$ to be $0.1$ and $20$, respectively.

This metric effectively measures the distance between two modes normalized by their variances, with a penalty when one mode is not highly weighted. These choices were made in order for the score to match by eye which distributions appear to have one as opposed to two peaks, and was found to empirically be more reliable than other statistics like peak-to-trough ratios or mode overlap. As demonstrated in Section~\ref{ssec:full_sim_suite}, a score of $\mathcal{B}>2.25$ appears to select the bimodal populations.

\section{Results}\label{sec:results}
\subsection{Surface Density Projections}\label{ssec:projections}
\begin{figure*}
  \centering

    \gridline{\fig{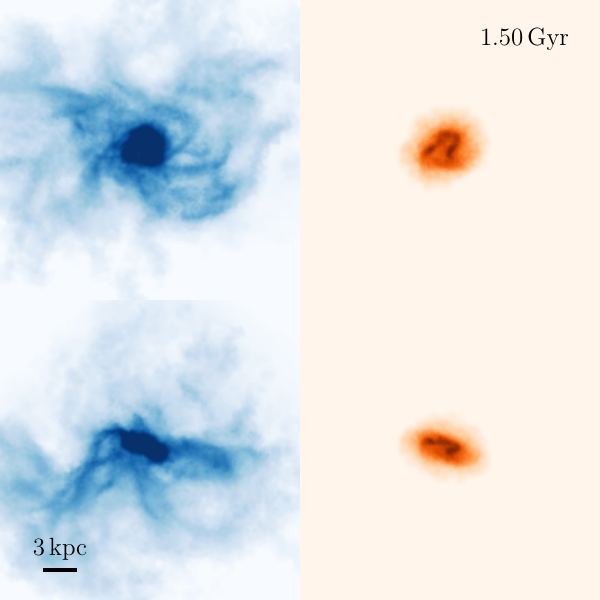}{0.3\textwidth}{}
              \fig{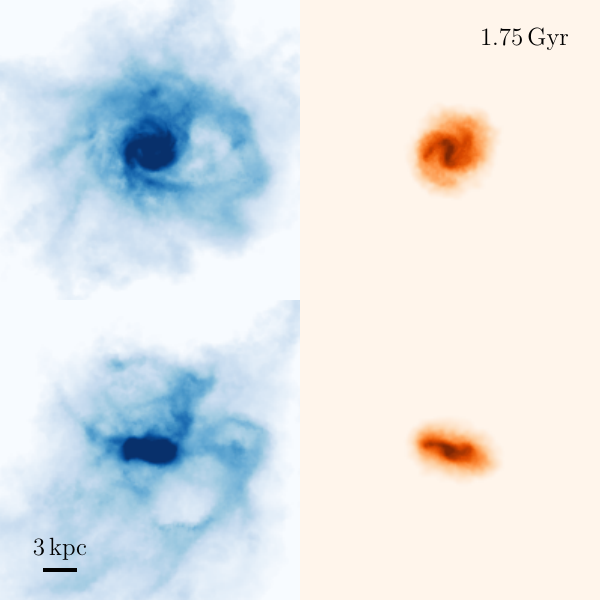}{0.3\textwidth}{}
              \fig{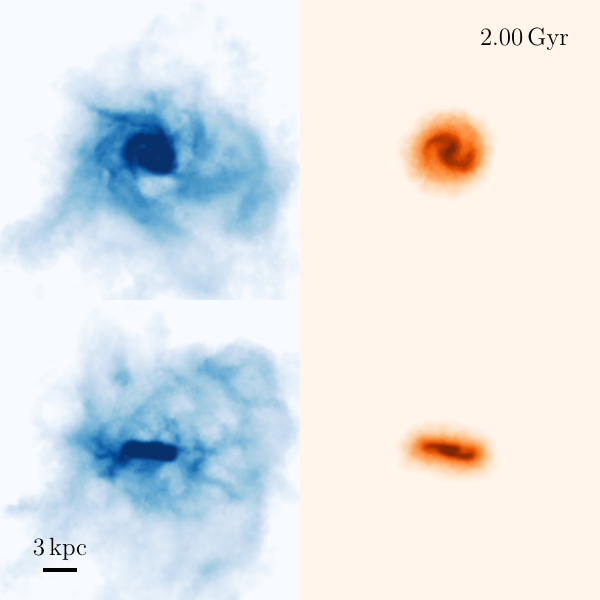}{0.3\textwidth}{}
              }
    \gridline{\fig{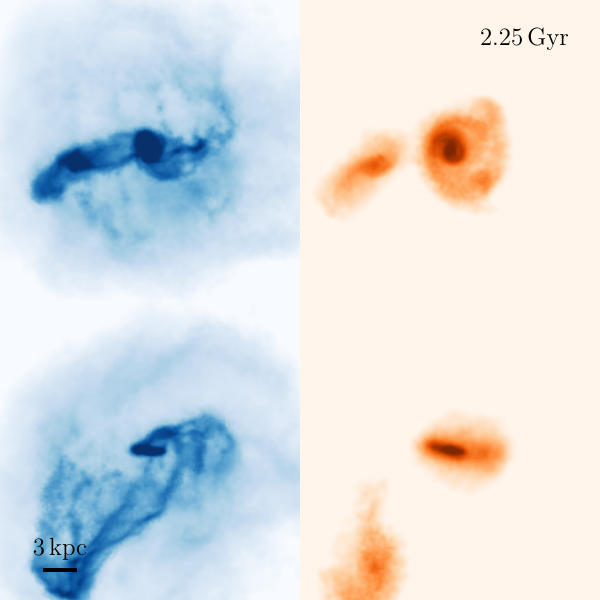}{0.3\textwidth}{}
              \fig{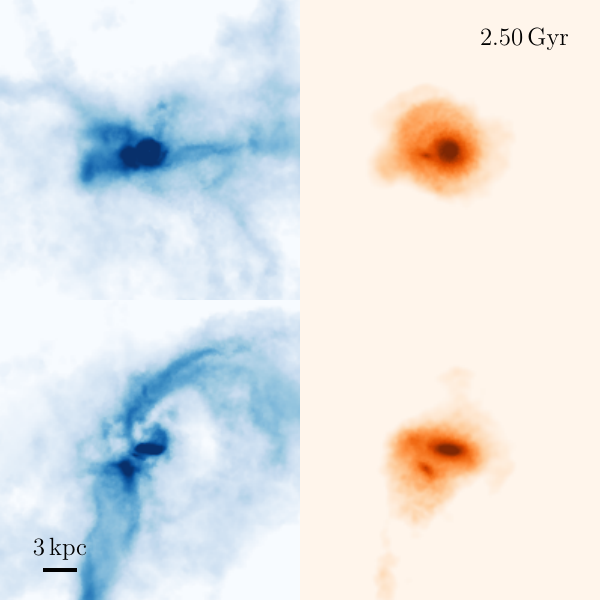}{0.3\textwidth}{}
              \fig{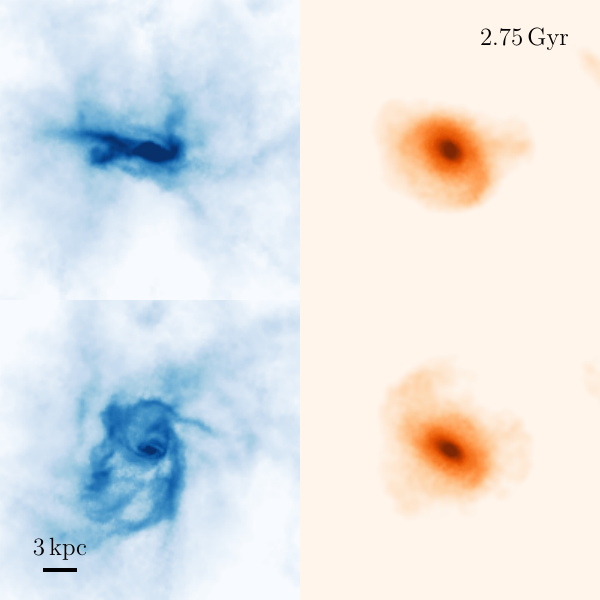}{0.3\textwidth}{}
              }
    \gridline{\fig{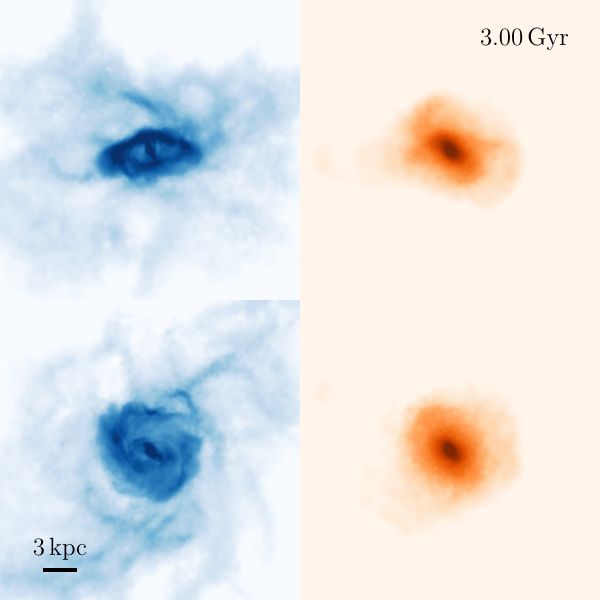}{0.3\textwidth}{}
              \fig{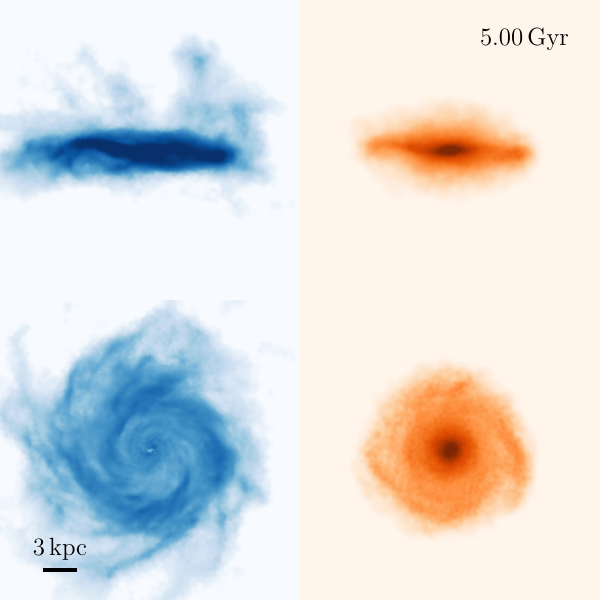}{0.3\textwidth}{}
              \fig{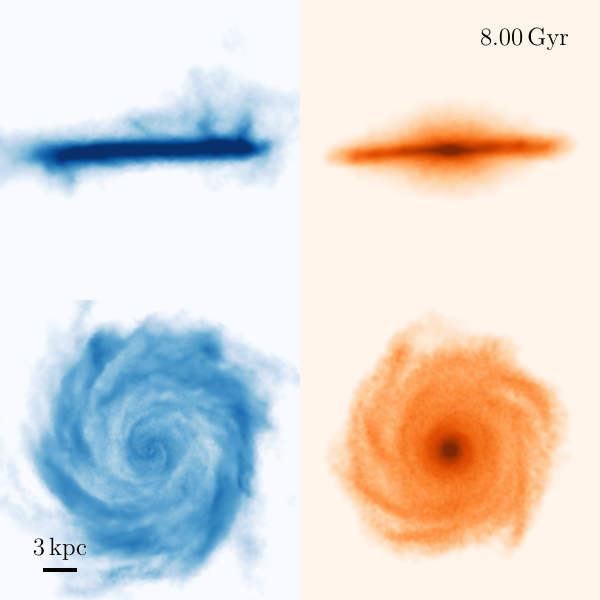}{0.3\textwidth}{}
              }
  \caption{Frames from a movie showing a surface density projection of the bimodal simulation over time. In each frame, the left/right (blue/orange) column shows the gas/star surface density. The upper/lower panels show the edge-on and face-on view. Every panel is oriented with respect to the final ($t=8\,\Gyr$) snapshot. The side-length of each panel is $30\,\kpc$, and the image is a projection through a box with the same side-length. The color map for the gas ranges from $1$ to $10^2\,\Msun/\pc^2$, while for the stars ranges from $1$ to $10^4\,\Msun/\pc^2$. Description of movie: The disk collapses quickly, forming a disk within $500\,\Myr$. The satellite galaxy quickly passes in the background at $\sim700\,\Myr$. At $\sim2\,\Gyr$, the satellite directly merges with the central galaxy, fully coalescing by $\sim3\,\Gyr$. Over the next $5\,\Gyr$, the disk steadily grows, expanding in size. (A full movie is available in published version.)}
  \label{fig:projections}
\end{figure*}

Before dissecting the simulated galaxies in detail, we first examine the surface density projections of the gas and stars in the central region, situated on the central galaxy in the bimodal simulation. This is shown in Figure~\ref{fig:projections}. In each grouping of four, the bottom and top rows show the face-on and edge-on views, respectively. The left and right columns show the gas (blue) and stellar (orange) surface density, respectively. Each panel is oriented with respect to the stellar angular momentum of the final snapshot ($t=8\,\Gyr$), computed using all star particles within the half-mass radius.

We can see that the galaxy grows in size after the merger. Additionally, the galaxy's orientation continues to change even within the last $3\,\Gyr$. The final galaxy is oriented $\sim126\degree$ with respect to the initial orientation of the central galaxy's gas halo. The dynamical and kinematic consequences of such a gas-rich merger is beyond the scope of our current work.

\subsection{Abundance Distribution}\label{ssec:abundplane}
\begin{figure*}
  \centering
  \includegraphics[width=\textwidth]{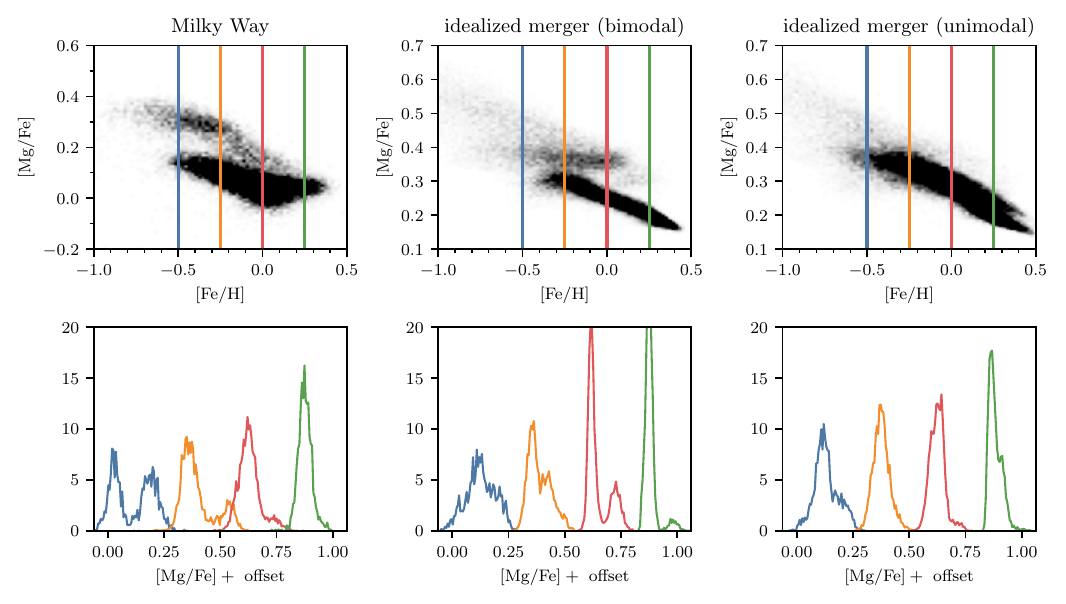}
  \caption{The abundance bimodality seen in the Milky Way can be reproduced in some idealized merger simulations. In the upper panels, we show the distribution of stars in the \MgFe{}-\FeH{} plane. The lower panels show the distribution of \MgFe{} at a fixed \FeH{} bin of width $0.05\,\dex$. The colors indicate the fixed \FeH values, which are $-0.5$, $-0.25$, $0$, and $0.25$. The left column shows the observed distribution in the Milky Way from ASPCAP DR17 \citep[][J.A.~Holtzman et al., in preparation]{2016AJ....151..144G}, while the right two columns show two idealized merger simulations. The idealized merger simulations are nearly identical, except that in the bimodal simulation the satellite has a starting velocity of $142\,\kms$, while in the unimodal simulation it has a starting velocity of $116\,\kms$. The labels ``unimodal'' and ``bimodal'' are of the \textit{outcome} of the simulation, and do not reflect a particular choice in the setup. The Milky Way (left column) exhibits a strong bimodal distribution of \MgFe{} at various \FeH{}. The idealized merger simulation marked as bimodal (center column) also exhibits a bimodal distribution of \MgFe{}, though the structure is not as strongly defined. The idealized merger simulation marked as unimodal (right column) exhibits only weak structure, if any at all.}
  \label{fig:fig1}
\end{figure*}

In Figure~\ref{fig:fig1}, we show the abundance distribution of the Milky Way as well as two of our idealized merger simulations in the upper panels. A number of our idealized simulations exhibit either a bimodal or unimodal abundance distribution, and so we have selected two representative examples as discussed in Section~\ref{ssec:orbit_setup}. The bimodal and unimodal labels are of the outcome of the simulation, and do not reflect any particular choice made in their setup.

There are, of course, differences between the bimodal simulation and the Milky Way. First, the scaling in \MgFe{} is different -- in the simulation, the low-$\alpha$ sequence lies at $\sim0.2$, while in the Milky Way it is at about $\MgFe\sim0$. Second, in the Milky Way the high-$\alpha$ sequence neatly joins the low-$\alpha$ sequence at $\FeH\sim0$, while in the simulation the two actually diverge more at higher \FeH{}.

In the lower panels of Figure~\ref{fig:fig1}, we show the distributions of \MgFe{} at different fixed \FeH{}. The (blue, orange, red, green) lines show the \MgFe{} distribution at a \FeH{} of ($-0.5$, $-0.25$, $0$, $0.25$), in bins of width $0.05\,\dex$. The distributions of \MgFe{} are offset (but not rescaled) so that they do not overlap. Here, the bimodality seen in the Milky Way is quite striking at lower metallicities. The peaks are well-separated, by $\sim0.2\,\dex$. In the bimodal simulation, the distribution is still clearly bimodal, but the peaks are less well-separated, by $\sim0.1\,\dex$. In the unimodal simulation, there is a hint of some structure at $\FeH > 0.25$, but there is not a strong multimodal structure.

\subsection{Build up of the Abundance Plane}\label{ssec:abundplane_build}
\begin{figure*}
  \centering
  \includegraphics[width=\textwidth]{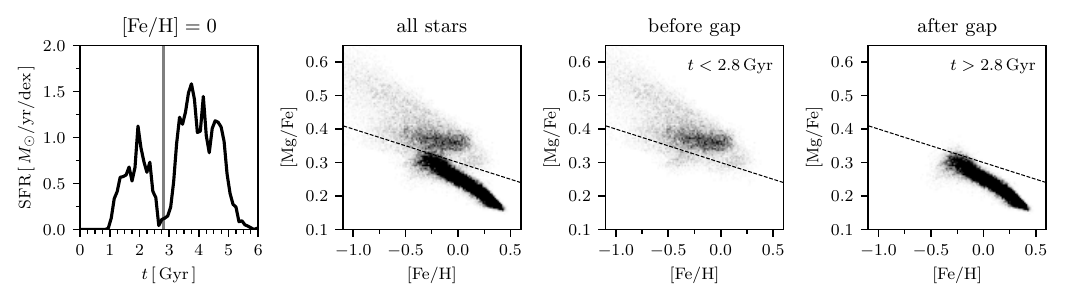}
  \caption{The buildup of the abundance plane in the bimodal simulation. The left panel shows the metallicity-dependent star formation rate (SFR) for star particles in the solar neighborhood at the end of the simulation, selected within a $0.1\,\dex$ bin centered at $\FeH=0$. A clear dip in this SFR occurs at $t \sim 2.8\,\Gyr$, marked by the vertical line. The center-left panel shows the abundance plane distribution for all stars in the solar neighborhood, with a dashed line at $-0.1\times\FeH+0.3$ (chosen by eye) demarcating the high- and low-$\alpha$ sequences. The center-right and right panels show the abundance plane distributions for stars formed before and after the SFR dip, respectively. The majority of the high-$\alpha$ sequence forms before the dip, while most of the low-$\alpha$ sequence forms afterward.}
  \label{fig:before_after}
\end{figure*}

Next, we examine the build up of the abundance plane. In the left panel of Figure~\ref{fig:before_after}, we show the metallicity-dependent star formation rate of the bimodal simulation using star particles in the solar neighborhood at the end of the simulation within a $0.1\,\dex$ bin centered at $\FeH=0$. There is a clear dip in the SFR at $\sim2.8\,\Gyr$, which we indicate with a vertical line.

In the center left panel, we show the abundance plane distribution from the bimodal simulation for all stars in the solar neighborhood, replicating Figure~\ref{fig:fig1}. A dashed line at $-0.1\times\FeH+0.3$, chosen by eye, is plotted to demarcate the high- and low-$\alpha$ sequences. In the center right and right panels, we show the distribution of star particles which form before and after the dip in the metallicity-dependent SFR at $t=2.8\,\Gyr$, respectively. The vast majority of the high-$\alpha$ sequence forms before the dip, while most of the low-$\alpha$ sequence forms afterward.

\begin{figure*}
  \centering
  \includegraphics[width=\textwidth]{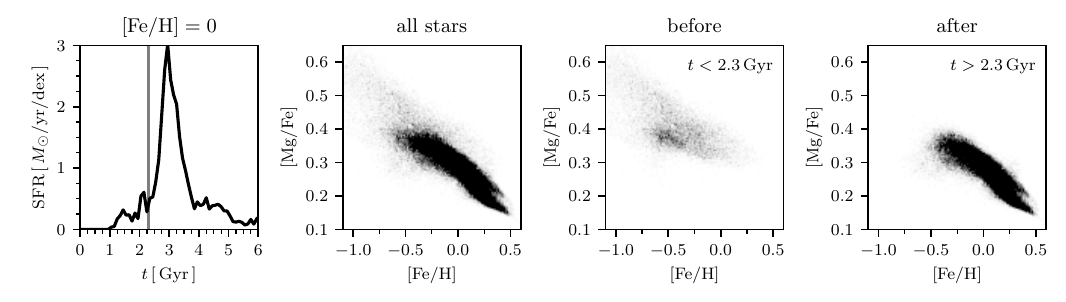}
  \caption{The buildup of the abundance plane in the unimodal simulation, similar to Figure~\ref{fig:before_after}. The left panel shows the metallicity-dependent star formation rate (SFR) for stars in the solar neighborhood, with no clear dip. A vertical line at $t=2.3\,\Gyr$, corresponding to the orbital stage of the gap in the bimodal simulation (see text), is included for comparison. The center left shows the abundance plane for all stars. The center-right and right panels show the abundance plane for stars formed before and after $t=2.3\,\Gyr$, respectively. Unlike the bimodal case, stars formed before and after this time overlap considerably in the abundance plane, indicating the absence of a distinct separation between high- and low-$\alpha$ sequences.}
  \label{fig:before_after_uni}
\end{figure*}

This sequence of build up is markedly different in the unimodal simulation, which we show in Figure~\ref{fig:before_after_uni}. Here, there is no clear dip in the metallicity dependent SFR in the left panel. We still place a vertical line similar to the one in Figure~\ref{fig:before_after}, except now it is chosen to be at $t=2.3\,\Gyr$, which is $\sim300\,\Myr$ after the second pericentric passage. This is where the gap appears in the bimodal simulation (see Figure~\ref{fig:orbit}). We see that stars which form before and after this point in the simulation overlap considerably in the abundance plane.

\begin{figure}
  \centering
  \includegraphics[width=\columnwidth]{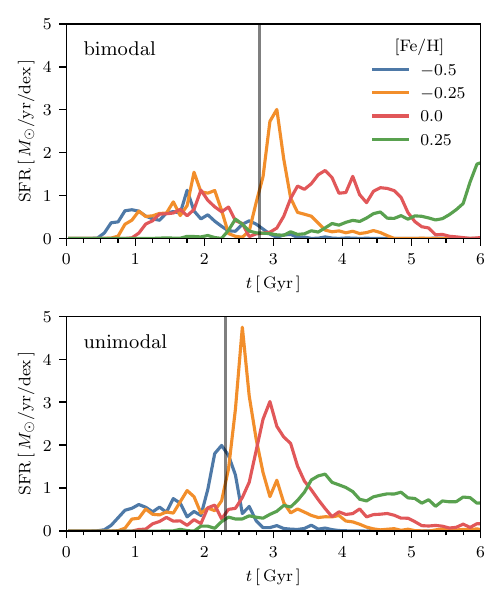}
  \caption{Metallicity-dependent star formation histories (SFH) for the bimodal (top) and unimodal (bottom) simulations. The SFH at different metallicities ($\FeH=-0.5, -0.25, 0.0, 0.25$) is shown in blue, orange, red, and green, respectively. In the top panel, the vertical line at $t=2.8\,\Gyr$ marks the SFR dip at $\FeH=0$, corresponding to the gap seen in Figure~\ref{fig:before_after}. A similar dip is present in the orange $\FeH=-0.25$ SFR, but it occurs $\sim250\,\Myr$ earlier, with a width of about $250\,\Myr$. The $\FeH=-0.5$ (blue) SFR lacks a distinct dip, while the $\FeH=0.25$ (green) SFR features an extended $\sim1\,\Gyr$ dip after a short period of SF, leading to a weak but well-separated high-$\alpha$ mode. The bottom panel shows the corresponding SFHs for the unimodal simulation, with a vertical line at $t=2.3\,\Gyr$. There are no strong dips at any metallicity, consistent with the absence of a well-defined bimodality.}
  \label{fig:before_after_sfh_by_iron}
\end{figure}

The timing and duration of the metallicity-dependent SFR dip is not the same for all \FeH{}. In the upper panel of Figure~\ref{fig:before_after_sfh_by_iron}, we show the SFR at metallicities of $-0.5$, $-0.25$, $0$, and $0.25$ in blue, orange, red, and green, respectively. We mark the location of the dip in the red $\FeH=0$ SFR with a vertical line at $t=2.8\,\Gyr$, as in Figure~\ref{fig:before_after}. In the orange SFR at $\FeH=-0.25$, which displayed a prominent bimodality in Figure~\ref{fig:fig1}, there is a similar dip. However, it occurs about $250\,\Myr$ earlier. The $\FeH=0$ and $\FeH=-0.25$ dips have widths of about $500$ and $250\,\Myr$, respectively. In the blue $\FeH=-0.25$ SFR, which does not display a prominent bimodality, there is no dip separating two periods of sustained SF. In the green $\FeH=0.25$ SFR, a small amount of SF occurs before an extended $\sim1\,\Gyr$ dip, leading to a bimodality with a weak although well-separated high-$\alpha$ mode.

In the lower panel of Figure~\ref{fig:before_after_sfh_by_iron}, we show corresponding curves for the unimodal simulation. A vertical line is shown at $2.3\,\Gyr$, as described earlier. Here, we see that at no \FeH{} is there a prominent dip.

\begin{figure*}
  \centering
  \includegraphics[width=\textwidth]{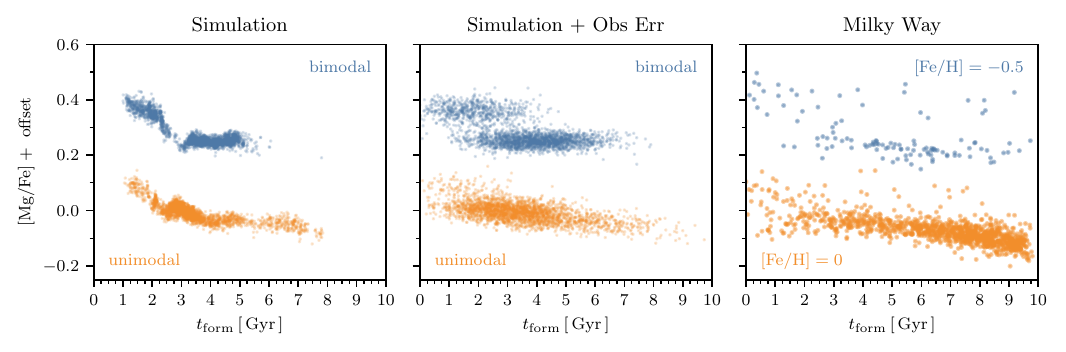}
  \caption{In the bimodal simulation, a gap appears when plotting stars in a $0.05\,\dex$ bin at $\FeH=0$ on the \MgFe{}-$t_{\textrm{form}}$ plane (blue, left panel). In the unimodal simulation there is no such gap (orange, left panel). In the middle panel, we show the same data but assuming errors in \FeH{}, \MgFe{}, and $t_{\textrm{form}}$ of $0.0075\,\dex$, $0.012\,\dex$, and $1\,\Gyr$, respectively. The gap in the bimodal simulation appears weakly as two populations overlapping in age. In the right panel we show stars in the Milky Way from APOKASC3 in $0.2\,\dex$ bins at $\FeH=-0.5$ (blue) and $0$ (orange). The low-metallicity bin is where the Milky Way bimodality is strongest while there is no bimodality at the solar-metallicity bin (see Figure~\ref{fig:fig1}). Offsets have been added to \MgFe{} values for clarity: $-0.3$ in the unimodal points in the left and middle panels, and $+0.1$ and $-0.1$ in the low/solar-metallicity bins, respectively, in the right panel.}
  \label{fig:alpha_vs_tform}
\end{figure*}

\subsection{Comparison to Observations}\label{ssec:compare_obs}
A plot of \MgFe{} vs age or formation time is a useful way to further demonstrate the formation scenario of the bimodality, as well as making comparison to observations. We show this for the simulation data in the left panel of Figure~\ref{fig:alpha_vs_tform} in bins of width $0.05\,\dex$ centered at $\FeH=0$. The bimodal simulation (blue, upper data) displays a gap in the ages at $\sim2.8\,\Gyr$, coinciding with the gap in Figure~\ref{fig:before_after}. On the other hand, the unimodal simulation (orange, lower data) displays no such age gap.

The center panel shows these simulation data points convolved with realistic observational errors. We assume errors in \FeH{}, \MgFe{}, and $t_{\textrm{form}}$ of $0.0075\,\dex$, $0.012\,\dex$, and $1\,\Gyr$, respectively.\footnote{The \FeH{} error impacts which star particles lie in the \FeH{} selection.} The \FeH{} and \MgFe{} errors are characteristic of the errors provided in the APOGEE dataset, and the $1\,\Gyr$ comes from assuming the typical $12.5\%$ uncertainty from APOKASC-3 at an age of $8\,\Gyr$. In the bimodal simulation (blue, upper data), separate populations can be seen that overlap significantly in $t_{\textrm{form}}$. There is not a strong separation in the unimodal case (orange, lower data).

The right panel shows observational data from APOKASC-3. We show in the blue, upper data a $0.2\,\dex$ bin centered at $\FeH=-0.5$ and in the orange, lower data centered at $\FeH=0$. In the Milky Way, there is a bimodal and unimodal \MgFe{} distribution at these metallicities, respectively. At $\FeH=-0.5$, one could weakly argue that there are separate populations as in the simulation. However, the large uncertainties at the relevant ages ($\sim1\,\Gyr$ at ages of $\sim8\,\Gyr$) and the small sample size (125 and 1175 at $\FeH=-0.5$ and $0$, respectively) prevents a definitive statement from being made. Larger sample sizes and more precise age estimates in the future would clarify the connection between simulation and observation, although achieving age uncertainties of $<10\%$ for such old stars is very challenging \citep[e.g.][]{2010ARA&A..48..581S}.

There is an additional complication in the data coming from the presence of young, $\alpha$-rich stars. These have been argued to be old stars with misclassified astroseismic ages due to binary mass transfer \citep[and references therein]{2023A&A...671A..21J}, though with some appearing to be genuinely young \citep[and references therein]{2024arXiv241002962L}, with a range of explanations given \citep[e.g.][]{2015A&A...576L..12C,2021MNRAS.508.4484J,2023arXiv231105815S}. It is unclear which of the $\alpha$-rich stars faithfully reflect the mean ISM chemistry at their inferred age, and so it is not obvious if any detailed conclusions can be drawn from this comparison.

\subsection{Full Simulation Suite}\label{ssec:full_sim_suite}
\begin{figure*}
  \centering
  \includegraphics[width=\textwidth]{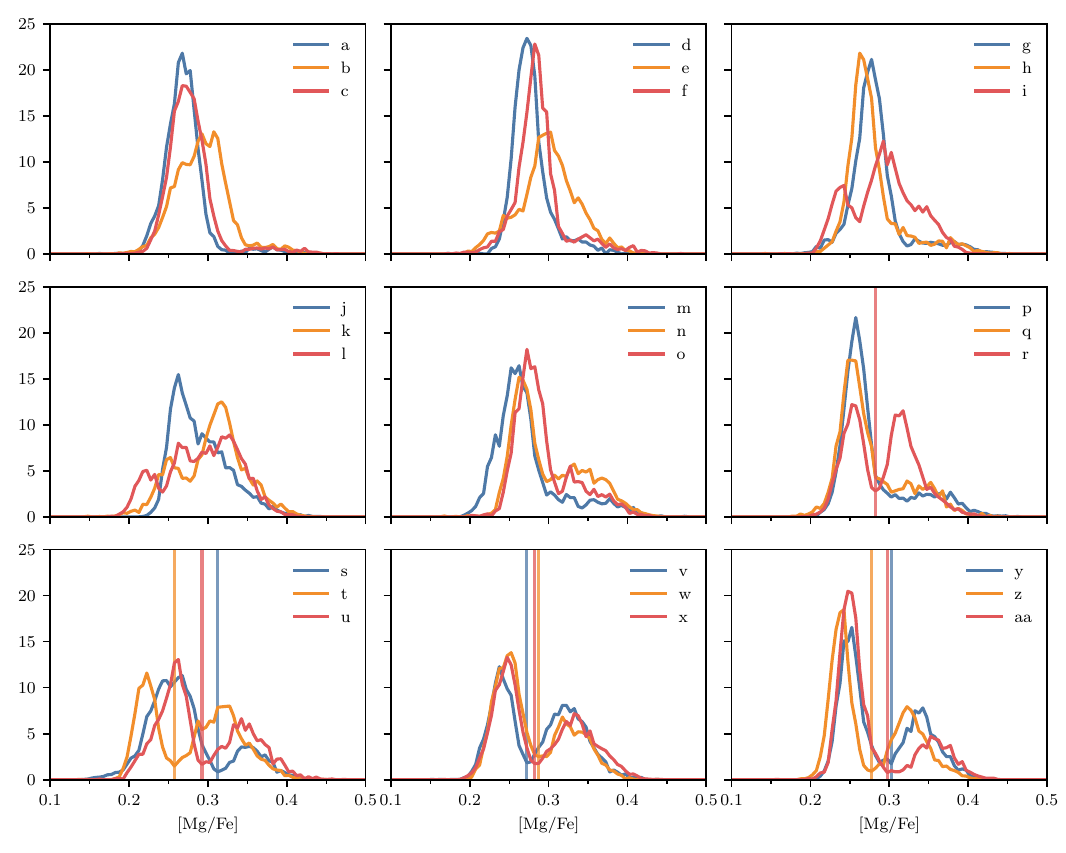}
  \caption{The conditional 1D \MgFe{} distribution for all stars with \FeH{} in a bin of width $0.1\,\dex$ centered at $0\,\dex$. Each panel corresponds to a different set of orbital parameters ($R_0$, $V_0$, and $\eta$) and is labeled alphabetically from least to most bimodal, with the labels defined in Table~\ref{tab:my-table}. The final simulation, which has the highest bimodality score $\mathcal{B}$, is labeled ``aa.'' Simulations with $\mathcal{B}>2.25$ (starting from simulation r) exhibit clear bimodalities, with the trough \MgFe{} (minimum between the two peaks) marked. This trough generally falls between $0.25$ and $0.3\,\dex$. Simulations with lower bimodality scores appear unimodal, though some (e.g., m, n, and o) show hints of a secondary population without forming a distinct bimodal structure.}
  \label{fig:all_hist}
\end{figure*}

We next expand our discussion to the full simulation suite by studying the conditional 1D \MgFe{} distribution in Figure~\ref{fig:all_hist}. We plot the distribution of \MgFe{} for all stars with \FeH{} lying in a bin of width $0.1\,\dex$ centered at $0\,\dex$. Each distribution corresponds to a choice in the three orbital variables $R_0$, $V_0$, and $\eta$. We rank the simulations in alphabetical order from least to most bimodal\footnote{There is one more simulation in the suite than letters in the English alphabet, so the simulation with the highest bimodality score is labeled aa.}, as determined by the bimodality score $\mathcal{B}$ introduced in Section~\ref{ssec:bim_score}. The list of simulations in the orbital grid along with the associated orbital parameters and bimodality scores is given in Table~\ref{tab:my-table}.

Simulations from r onward, which have $\mathcal{B}>2.25$, are considered bimodal. This demarcation was chosen by eye using Figure~\ref{fig:all_hist}. For these simulations, we plot the trough \MgFe{} (the location of the minimum between the two maxima) as a vertical line. This value was determined by taking the location of the minimum of the distribution between $0.25$ and $0.35\,\dex$, and is also given in Table~\ref{tab:my-table}.

Simulations from r to aa have clear bimodalities. The trough \MgFe{} is roughly consistent between simulations, appearing between $\sim0.25$ and $0.3\,\dex$. For simulations marked as unimodal, some appear to have two populations but which are not distinct enough to form a clear bimodality (e.g., m, n, and o).

\begin{table}[]
\centering
\begin{tabular}{cccccc}
$\textrm{letter}$ & $R_0$ & $V_0$ & $\eta$ & $\mathcal{B}$ & $\textrm{trough}\,\MgFe{}$ \\ \hline
a                 & $129$ & $129$ & $0.4$  & $0.29$        &                            \\
b                 & $129$ & $116$ & $0.4$  & $0.35$        &                            \\
c                 & $116$ & $116$ & $0.4$  & $0.77$        &                            \\
d                 & $142$ & $116$ & $0.5$  & $0.79$        &                            \\
e                 & $142$ & $129$ & $0.6$  & $0.89$        &                            \\
f                 & $116$ & $129$ & $0.4$  & $1.12$        &                            \\
g                 & $116$ & $116$ & $0.5$  & $1.25$        &                            \\
h                 & $116$ & $142$ & $0.4$  & $1.34$        &                            \\
i                 & $142$ & $142$ & $0.5$  & $1.43$        &                            \\
j                 & $129$ & $142$ & $0.5$  & $1.46$        &                            \\
k                 & $142$ & $129$ & $0.4$  & $1.62$        &                            \\
l                 & $142$ & $142$ & $0.6$  & $1.75$        &                            \\
m                 & $116$ & $129$ & $0.5$  & $1.82$        &                            \\
n                 & $129$ & $129$ & $0.5$  & $1.96$        &                            \\
o                 & $129$ & $116$ & $0.5$  & $2.04$        &                            \\
p                 & $116$ & $116$ & $0.6$  & $2.16$        &                            \\
q                 & $116$ & $129$ & $0.6$  & $2.24$        &                            \\ \hline
r                 & $142$ & $129$ & $0.5$  & $2.26$        & $0.283$                    \\
s                 & $142$ & $116$ & $0.4$  & $2.52$        & $0.313$                    \\
t                 & $116$ & $142$ & $0.6$  & $2.59$        & $0.258$                    \\
u                 & $116$ & $142$ & $0.5$  & $2.62$        & $0.293$                    \\
v                 & $142$ & $116$ & $0.6$  & $2.65$        & $0.273$                    \\
w                 & $129$ & $142$ & $0.6$  & $2.66$        & $0.288$                    \\
x                 & $142$ & $142$ & $0.4$  & $2.70$        & $0.283$                    \\
y                 & $129$ & $116$ & $0.6$  & $2.94$        & $0.303$                    \\
z                 & $129$ & $129$ & $0.6$  & $3.12$        & $0.278$                    \\
aa                & $129$ & $142$ & $0.4$  & $3.34$        & $0.298$                   
\end{tabular}
\caption{All simulations in the orbital grid (as defined by $R_0$, $V_0$, and $\eta$) ordered by their bimodality score $\mathcal{B}$. Each simulation is given an identifying letter. For simulations marked as bimodal ($\mathcal{B}>2.25$), we also list the trough \MgFe{}, or location of the minimum of the \MgFe{} distribution.}
\label{tab:my-table}
\end{table}

\begin{figure*}
  \centering
  \includegraphics[width=\textwidth]{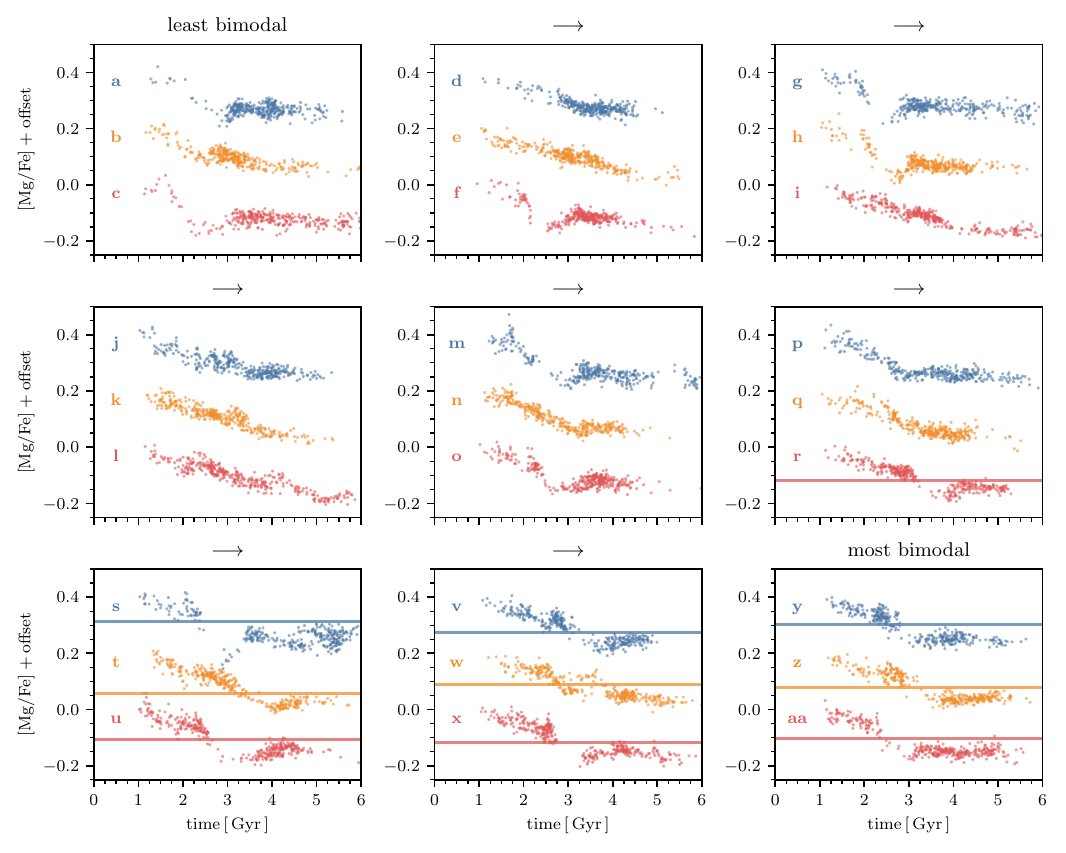}
  \caption{Scatter plot of \MgFe{} versus formation time for star particles in each simulation, corresponding to the 1D distributions shown in Figure~\ref{fig:all_hist}. Each panel represents a different set of orbital parameters ($R_0$, $V_0$, and $\eta$), arranged from least to most bimodal (as defined by the bimodal score $\mathcal{B}$. The colors and order match Figure~\ref{fig:all_hist}, with a random subsample of 350 stars plotted per simulation. Transparency ($\alpha=0.5$) is used to highlight regions of higher density. The second and third (orange and red) simulations in each panel are offset by $-0.2$ and $-0.4\,\dex$ for clarity, respectively. For bimodal simulations (r through aa), the trough \MgFe{} (minimum between the two peaks) is indicated with a horizontal line. In most bimodal cases, a gap in the distribution emerges at approximately the merger time ($\sim2$--$3,\Gyr$, depending on orbital parameters), with older stars forming at higher \MgFe{} and younger stars at lower \MgFe{}. Notably, simulation~t lacks a clear gap, and simulation~w exhibits irregular behavior with star formation occurring both above and below the trough. Among unimodal simulations (a through q), some exhibit apparent gaps (e.g., d, f, g, h, and o), but these do not result in strong bimodalities due to the low number of high-\MgFe{} stars forming before the gap.}
  \label{fig:all_scatter}
\end{figure*}

We study the formation of the 1D distributions in Figure~\ref{fig:all_hist} through a scatter plot of \MgFe{} vs formation time of star particles in Figure~\ref{fig:all_scatter}, similar to Figure~\ref{fig:alpha_vs_tform}. The order and colors are identical as in Figure~\ref{fig:all_hist}, and we use the same \FeH{} selection. We only plot a random subsample of 350 stars, and points are plotted with a transparency of $\alpha=0.5$ so that the perceived density is slightly higher in cases of overlap. An offset of $-0.2$ and $-0.4$ are given to the second and third (orange and red) simulations in each panel. For simulations marked as bimodal (r through aa), we also plot a horizontal line at the location of the trough \MgFe{}.

Most of the bimodal simulations (r through aa) have a gap in the distribution at around the time of the merger ($\sim2$-$3\,\Gyr$, depending on the orbital parameters). Before this gap, star formation occurs above the trough while after it occurs below the trough. However, there are two exceptions: simulation~t does not have a full gap (though the density does decrease), and simulation~w is irregular, with star formation simultaneously occurring above and below the trough.

In the unimodal simulations (a through q), there does appear to be gaps (d, f, g, h, and o). However, in these cases, there is not a large amount of star formation occurring before the gap, and so the high-$\alpha$ sequence is underemphasized and does not lead to a significant bimodality.

\section{Discussion}\label{sec:discussion}
We have investigated the formation of an $\alpha$-element bimodality in the Milky Way through a series of idealized merger simulations. Our key finding is that a metallicity-dependent quiescent period in star formation can lead to a bimodal distribution in \alphaFe{} at specific \FeH{} values. This scenario does not necessarily require a global quenching period. We now discuss the details of this mechanism, its connection to high-redshift observations and cosmological simulations, compare with some explanations in the literature, and explore its observational implications and directions for future work.

\subsection{\FeH{}-dependent Quiescence Leads to Bimodality}\label{ssec:formqui}
We executed a series of idealized merger simulations in which we modified the starting radius and velocity by $\pm10\%$ and the circularity by $\pm0.1$, for a total of $27$ simulations. The central and satellite galaxies, which are meant to resemble the Milky Way and GSE at $z\sim2$, are otherwise identical across the simulations. Some of these simulations induce a bimodality, while others do not. We have examined a representative of each scenario in detail.

The key driver of bimodality in our simulations is the presence of a metallicity-dependent quiescent period. This is shown most clearly in Figure~\ref{fig:before_after}, where the abundance plane is split into stars which form before and after a gap in the $\FeH=0$ SFR. Stars which form before and after the gap populate the high- and low-$\alpha$ sequences, respectively, with minimal overlap. No gap and no separation between the sequences is seen in the unimodal case (Figure~\ref{fig:before_after_uni}).

This perspective is bolstered by examining the distributions of \MgFe{} at $\FeH=0$ for the full simulation suite (Figure~\ref{fig:all_hist}). We order these simulations alphabetically using the bimodality score ($\mathcal{B}$, see Section~\ref{ssec:bim_score}). Simulations~r through aa are considered bimodal based on a visual inspection, and the location of their trough \MgFe{} (minimum of the distribution) is indicated with a vertical line.

The formation of these bimodal populations can be understood from Figure~\ref{fig:all_scatter}, which shows a scatter plot of \MgFe{} and formation time in the same order, with the trough \MgFe{} indicated with a horizontal line. Here, we can see that bimodal simulations tend to have a gap shortly after their respective mergers ($\sim2$--$3\,\Gyr$, depending on the orbital configuration), lying at the position of the trough \MgFe{}.

There are two exceptions in the bimodal cases. First, simulation~t does not have a complete gap, although the number of stars forming does still drop. This indicates that the complete absence of star formation is not necessary, but rather a reduction in the \FeH{}-dependent SFR may be sufficient if the \alphaFe{} ratio is declining fast enough. Second, simulation~w exhibits some irregular behavior, with star formation switching between high- and low-$\alpha$ multiple times.

There are a few unimodal simulations that have gaps -- e.g., d, f, g, h, and o. However, in these cases there is very little star formation at $\FeH=0$ before the gap, and so in these cases there is not a distinct high-$\alpha$ mode that can form.

Overall, these simulations indicate that a gap in star formation at a specific metallicity can lead to a bimodality in the conditional \alphaFe{} distribution at that metallicity. Determining precisely the mechanism behind generating these gaps is beyond the scope of this work, but we speculate briefly in Section~\ref{ssec:future_work}.

\subsection{Connection to High Redshift Quenching}\label{ssec:obshiz}
One plausible avenue to producing a \FeH{}-dependent halt in star formation is through a global quiescent period (see Appendix~\ref{app:all_sfh}). Galaxies which undergo a starburst to quiescence to rejuvenation sequence (post-starburst galaxies, or PSBs) are observed at high-$z$ ($z>1$), and may be plausible Milky Way-progenitors. With abundance matching, we expect the Milky Way's total stellar mass to be $\sim10^{10.3}\,\Msun$ at $z\sim2$ \citep{2013ApJ...771L..35V}. A number of authors have explored PSBs and quiescent galaxies at slightly higher masses at $z\sim2$, with large advances in the post-JWST era.

First, PSBs are not uncommon. \citet{2023ApJ...953..119P} found that in massive galaxies ($M_* > 10^{10.6}\,\Msun$) the fraction of PSBs (inferred ages $< 800\,\Myr$) increases from $\sim2.7\%$ ($99/3655$) at $1.0 < z < 1.44$ to $\sim8\%$ ($89/1118$) at $2.16 < z < 2.5$ \citep[see also][]{2012ApJ...745..179W,2019ApJ...874...17B}. Later, \citet{2024arXiv240417945P} found that $\sim10\%$ of galaxies at $\sim10^{10.3}\,\Msun$ are quenched \citep[consistent with][]{2013ApJ...777...18M}, and $\sim30\%$ of their quiescent sample is a PSB at $z\sim2$. If these galaxies can be quickly rejuvenated, as the system studied in this work would suggest, then the total fraction of galaxies that go through a starburst-quenching phase may be higher.

Furthermore, \citet{2023arXiv231215012C} found that lower mass quiescent galaxies (towards $10^{10.3}\,\Msun$) tend to be younger and more disky, pointing to a merger driven scenario. There is also evidence that AGN, which we suspect might be responsible for the star formation gaps in our system (Appendix~\ref{app:cause_qui}), is operating at these redshifts \citep[e.g.][and references therein]{2023arXiv230806317D,2024arXiv240417945P,2024arXiv240518685M,2024Natur.630...54B}.

In the context of our proposed mechanism, only a metallicity-specific quenching period is necessary for generating an $\alpha$-bimodality. This may correspond to inside-out or outside-in quenching, for which examples are known in the local and high-$z$ universe \citep[e.g.,][]{2015Sci...348..314T,2019ApJ...872...50L}.

\subsection{Connection to Cosmological Simulations}\label{ssec:cosmo}
As discussed in Section~\ref{sec:intro}, several authors have examined the formation of abundance plane structure in cosmological simulations. Of most interest to us is the zoom Au~23 in \citet{2018MNRAS.474.3629G}. This galaxy, one of six considered in their work, exhibits a bimodality that extends beyond the inner disk. The interpretation given by the authors is of a ``shrinking'' gaseous disk. This is equivalent to saying that the outer disk becomes depleted of gas. This shrinking of the disk, which occurs at $t_{\textrm{lookback}}\sim6\,\Gyr$, is associated with a dip in the SFR at that radius and a decrement in the median \alphaFe{} of $\sim0.05\,\dex$ (their Figure~2), which shortly after recovers. This sequence of events is more extended than in our work, but it resembles the scenario in Figure~\ref{fig:before_after}.

\citet{2018MNRAS.477.5072M} found that Milky Way-like bimodalities are rare in EAGLE, occurring in $\sim5\%$ of galaxies. \citet{2021MNRAS.501..236D,2022MNRAS.515.1430D} showed that merger-induced quenching in zooms can occur in the EAGLE model \citep[see also][]{2017MNRAS.465..547P}. However, the situation may be different in the lower resolution large box. Furthermore, if the proposed starburst-quenching phase is driven by AGN feedback, then the outcome of any particular cosmological simulation with regards to the bimodality is intimately tied to its AGN model. Unfortunately, such models are highly uncertain \citep[e.g.][]{2022MNRAS.511.3751H}.

\citet{2023arXiv231016085K} explored the impact of quenching in the Magneticum Pathfinder suite. They found that galaxies which quench undergo a starburst followed by an AGN-driven quenching phase. In the post-starburst regime, they claim galaxies are $\alpha$-enhanced. We do find that the bulk stellar \MgFe{} is enhanced after the merger in our bimodal simulation compared to the isolated simulation, but only at the $\sim0.01\,\dex$ level.

\subsection{Infall Interpretations}\label{ssec:dilute}
In some previous work, it was reported that the bimodality is a consequence of a sudden deposition of metal-poor, $\alpha$-poor gas by a satellite or cosmological filament -- i.e., a ``dilution'' \citep{2020MNRAS.491.5435B,2021MNRAS.503.5846R}. The separation of the sequences follows from the rapidity of the dilution. This was elaborated upon by \citet{2021MNRAS.503.5868R} who described a zoom where the low-$\alpha$ disk forms out of a relatively pristine cosmological filament. This disk is inclined relative to the high-$\alpha$ disk, with the two disks later tidally realigning. The longer-standing two-infall class of models argue that the two sequences diverge due to two episodic accretion episodes, with some possible enrichment of the second episode arising from an associated satellite \citep{1997ApJ...477..765C,2009IAUS..254..191C,2017MNRAS.472.3637G,2019A&A...623A..60S}.

We have shown that minor changes to the orbit of our idealized merger can result in outcomes that are either bimodal or unimodal. The content of gas that is delivered to the system is nearly identical regardless of the orbit, and so the dilution interpretation is not applicable to our simulations. That being said, a removal of gas from the system either through star formation or through ejection could make dilution from infalling gas more efficient, so the two scenarios are not mutually exclusive.

It was recently elaborated by \citet{2024arXiv240511025S} that these models also argue for a star formation gap between the two accretion episodes \citep[see also][]{1996ASPC...92..307G,1998A&A...338..161F,2000A&A...358..671G,2015A&A...578A..87S,2020A&A...640A..81N}. This gap is starvation-driven and can last several \Gyr{}. The present work argues for a starburst-driven quiescence followed by a rapid rejuvenation, with the entire process taking less than $1\,\Gyr$ and the gap only lasting a few hundred \Myr{}. The physical origin and some details are different, but one can appreciate that the bimodality arises from a similar process.\footnote{Compare Section~\ref{ssec:formqui} to the first key result in the Conclusions of \citet{2024arXiv240511025S}.}

\subsection{Direct Observational Test}\label{ssec:obsqui}
Figure~\ref{fig:before_after_sfh_by_iron} indicates a very direct observational test of the mechanism proposed in this work: for disk stars at a given \FeH{}, there should be a gap of $\sim300\,\Myr$ in ages at $\sim8\,\Gyr$, though in the Milky Way the gap could be larger. With a survey of properly chosen stars, this gap could be directly measured with a modestly sized (few hundred) sample of old stars with age uncertainties of a few percent. To our knowledge, the best method at these ages is differential analysis of solar twins, which can provide an age uncertainty of $\sim5\%$ \citep[e.g.][]{2014ApJ...795...23B,2018MNRAS.474.2580S}. However, this has only been applied to stars with solar metallicity, where there is not a clear separation between the high- and low-$\alpha$ sequences (though a gap in ages may still be present). A sort of differential approach could be applied also at lower metallicities, which would simply lack the absolute age calibration that the Sun provides.

The gap could also be indirectly probed by a larger sample of slightly less precise ages. Astroseismology appears to be the most promising avenue. Currently the largest sample is APOKASC-3 \citep{2025ApJS..276...69P}, which has $\sim2\textrm{k}$ stars with ages measured to $<12.5\%$ precision.\footnote{Errors here taken to be the maximum of the upper and lower age estimates in the APOKASC-3 catalog.} The upcoming PLATO mission is looking to measure $\sim20\textrm{k}$ stars with ages measured to $<10\%$ precision \citep{2024arXiv240605447R}. Further work is needed to determine how well these surveys could constrain the gap.

\subsection{Future Work}\label{ssec:future_work}
The largest unanswered question in the current work is the origin of gaps in the metallicity-dependent SFR. In Appendix~\ref{app:cause_qui}, we show that the merger is associated with strong feedback from the central AGN. Determining the precise mechanism and relationship between the two is delayed to future work. Furthermore, other mechanisms of quenching might be able to produce the gaps studied here, and so their exploration is worthwhile.

Another natural next step would be to extend the idealized simulations in this work to a wider range of orbits, galaxy properties, and feedback models. However, several aspects of our setup are unrealistic, for example, (1) the simulation is not in an expanding universe, (2) our feedback model is weaker than typical ones calibrated to full cosmological box simulations, (3) the initial conditions have a steeper potential well prior to star formation than in the real universe, and (4) our halos lack small-scale substructure. The simplicity of our setup aids interpretation, but also limits it applicability to the real universe. Something along the lines of the genetic modification technique to explore various mergers as done in this work may be useful \citep{2016MNRAS.455..974R,2017MNRAS.465..547P}.

There is, of course, still great uncertainties in the stellar evolution models commonly adopted by different groups. Initial work on systematically exploring the stellar evolution parameters has been done by \citet{2017A&A...605A..59R,2021MNRAS.508.3365B}. Exploring these variations in the simulations presented in this work would be interesting, though exploring their interactions with brief quiescent periods in simpler chemical evolution models may be a better first step.

There is also the perennial problem of diffusion within the hydrodynamics solver. In purely Lagrangian solvers, there is no diffusion between resolution elements, while in Eulerian codes the diffusion can be quite high.\footnote{No galaxy formation simulation is fully Lagrangian since there must be, at a minimum, mass exchange between star particles and gas. Here we just mean that there is no mass exchange between gas elements.} AREPO limits the numerical diffusion by allowing the mesh to move in a quasi-Lagrangian manner, and using a second-order solver \citep{2010MNRAS.401..791S}. In FIRE-2 \citep{2018MNRAS.480..800H}, which uses the Lagrangian code GIZMO \citep{2015MNRAS.450...53H}, a subgrid turbulent metal diffusion model was used. It would be interesting to see how models with different diffusivity properties would relax or strengthen the necessity of a quiescent period to produce a bimodality.

\section{Conclusion}\label{sec:conclusion}
The \alphaFe{}-\FeH{} plane of stellar abundances is a record of the gas-phase abundances of the Galaxy. In this plane, a bimodality has now been definitively measured. Proposals for its formation include radial migration, particular gas infall scenarios, and galaxy mergers.

In this work, we have shown that a brief ($\sim300\,\Myr$) period of halted star formation in a narrow \FeH{} bin is capable of producing a bimodal distribution in \alphaFe{} at that \FeH{}. This proposal requires that the \alphaFe{} of gas within the galaxy is decreasing with time so that the gap in star formation translates to a gap in \alphaFe{}. A global quiescent period can satisfy these constraints, but is not necessary. We demonstrate the plausibility of this scenario using a grid of idealized merger simulations with slightly varied orbital parameters. This scenario could potentially be triggered in non-merger scenarios, which we plan to explore in future work.

This scenario leads to the natural prediction that for stars occupying a narrow bin in \FeH{} where the bimodality is present ($\FeH\lesssim-0.2$), there should be a gap in ages for $\sim8\,\Gyr$ old stars with a width of $\sim300\,\Myr$. Currently, the best age estimates for such stars have errors of $\sim1\,\Gyr$. However, future observations targeting \textit{relative} ages of such stars might be able to achieve the necessary precision. Our proposed mechanism may operate in many external galaxies, whether or not these \FeH{}-dependent metallicity gaps are merger-induced.

\begin{acknowledgements}
We would like to thank the anonymous referee for their helpful comments which have served to clarify and strengthen our main arguments. We would like to thank Megan Bedell, Christopher Carr, Vedant Chandra, Charlie Conroy, Drummond B. Fielding, Lars Hernquist, Federico Marinacci, Rohan P. Naidu, Melissa K. Ness, Minjung Park, Vadim A. Semenov, and Turner Woody for helpful discussions. We would like to thank Marc Pinsonneault for sharing early access to the APOKASC-3 dataset. We would like to thank Filippo Barbani for kindly providing his version of \texttt{MakeNewDisk}. A.B. would like to thank Todd Phillips for helpful discussions.

This work has made use of data from the European Space Agency (ESA) mission {\it Gaia} (\url{https://www.cosmos.esa.int/gaia}), processed by the {\it Gaia} Data Processing and Analysis Consortium (DPAC, \url{https://www.cosmos.esa.int/web/gaia/dpac/consortium}). Funding for the DPAC has been provided by national institutions, in particular the institutions participating in the {\it Gaia} Multilateral Agreement.

Funding for the Sloan Digital Sky 
Survey IV has been provided by the 
Alfred P. Sloan Foundation, the U.S. 
Department of Energy Office of 
Science, and the Participating 
Institutions. 

SDSS-IV acknowledges support and 
resources from the Center for High 
Performance Computing  at the 
University of Utah. The SDSS 
website is www.sdss4.org.

SDSS-IV is managed by the 
Astrophysical Research Consortium 
for the Participating Institutions 
of the SDSS Collaboration including 
the Brazilian Participation Group, 
the Carnegie Institution for Science, 
Carnegie Mellon University, Center for 
Astrophysics | Harvard \& 
Smithsonian, the Chilean Participation 
Group, the French Participation Group, 
Instituto de Astrof\'isica de 
Canarias, The Johns Hopkins 
University, Kavli Institute for the 
Physics and Mathematics of the 
Universe (IPMU) / University of 
Tokyo, the Korean Participation Group, 
Lawrence Berkeley National Laboratory, 
Leibniz Institut f\"ur Astrophysik 
Potsdam (AIP),  Max-Planck-Institut 
f\"ur Astronomie (MPIA Heidelberg), 
Max-Planck-Institut f\"ur 
Astrophysik (MPA Garching), 
Max-Planck-Institut f\"ur 
Extraterrestrische Physik (MPE), 
National Astronomical Observatories of 
China, New Mexico State University, 
New York University, University of 
Notre Dame, Observat\'ario 
Nacional / MCTI, The Ohio State 
University, Pennsylvania State 
University, Shanghai 
Astronomical Observatory, United 
Kingdom Participation Group, 
Universidad Nacional Aut\'onoma 
de M\'exico, University of Arizona, 
University of Colorado Boulder, 
University of Oxford, University of 
Portsmouth, University of Utah, 
University of Virginia, University 
of Washington, University of 
Wisconsin, Vanderbilt University, 
and Yale University.

We acknowledge the use of OpenAI’s ChatGPT and Anthropic's Claude for assistance in editing this manuscript for clarity and conciseness and in generating small analysis scripts and code snippets.

\end{acknowledgements}

\bibliography{ref}{}
\bibliographystyle{aasjournal}

\software{ {\sc astropy} \citep{astropy:2013,astropy:2018,astropy:2022}, {\sc h5py} \url{http://www.h5py.org/}, {\sc inspector\_gadget} \url{https://bitbucket.org/abauer/inspector_gadget/}, {\sc joblib} \url{https://joblib.readthedocs.io/en/latest/}, {\sc matplotlib} \citep{Hunter:2007}, {\sc numba} \citep{lam2015numba}, {\sc numpy} \citep{harris2020array}, {\sc scikit-learn} \citep{scikit-learn}, {\sc scipy} \citep{2020SciPy-NMeth}, {\sc tqdm} \url{https://tqdm.github.io/}, {\sc vortrace} \url{https://github.com/gusbeane/vortrace}}

\appendix
\section{Star Formation Histories}\label{app:all_sfh}
\begin{figure*}
  \centering
  \includegraphics[width=\textwidth]{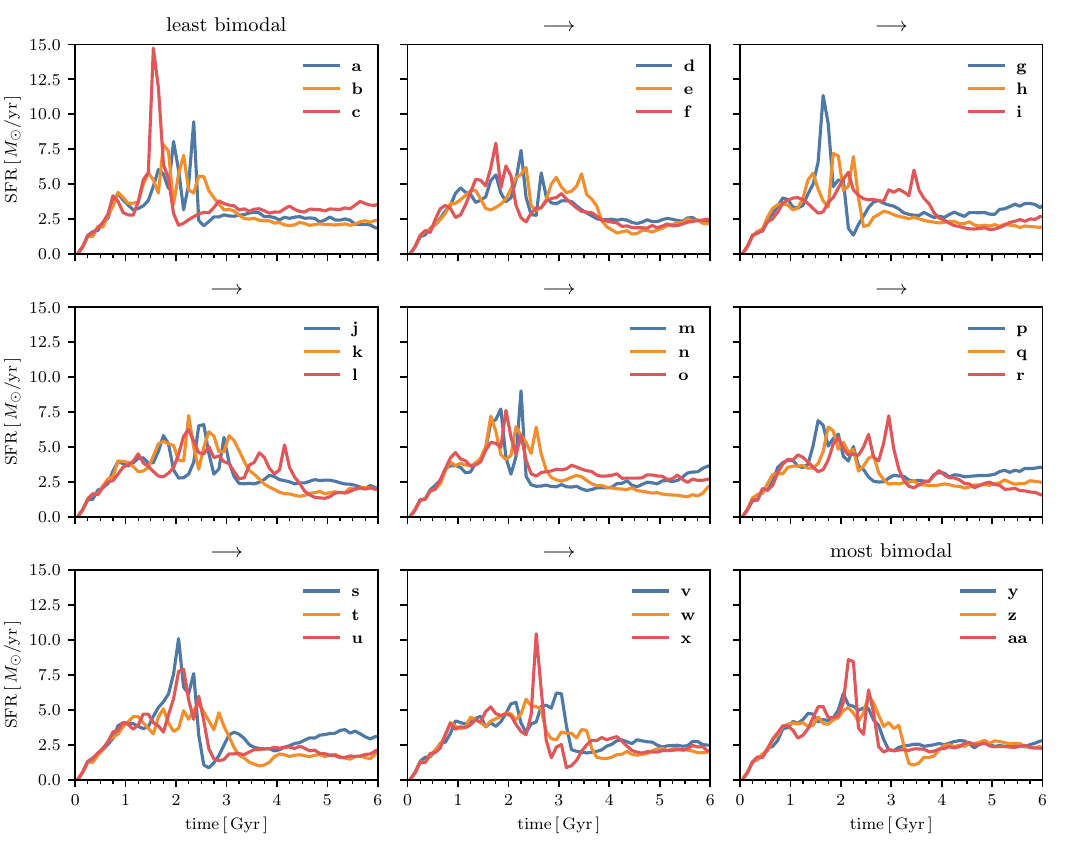}
  \caption{Global star formation history (SFH) for each simulation, plotted in the same order as Figures~\ref{fig:all_hist} and \ref{fig:all_scatter}, with increasing bimodality score $\mathcal{B}$ from left to right. The colors correspond to the same simulations as in previous figures. A global quiescent period, characterized by a significant dip in the SFR, is observed in simulations~g, s, and x. Among these, simulations~s and x exhibit strong bimodal \MgFe{} distributions, while simulation~g remains unimodal due to insufficient early star formation at $\FeH\sim0$ before the quiescent phase. Most other simulations do not display a clear global quiescent period, indicating that such a phase is not strictly necessary for bimodality to emerge.}
  \label{fig:all_sfh}
\end{figure*}

One potential avenue for creating an \FeH{}-dependent star formation gap is through quiescence. This is demonstrated by examining the global SFH in Figure~\ref{fig:all_sfh}, with the panels and colors showing each simulation in the grid ordered by their bimodality score $\mathcal{B}$ in the same way as Figures~\ref{fig:all_hist} and \ref{fig:all_scatter}. One can see that there is a global quiescent period in simulations~g, s, and x. Simulations~s and x have a bimodal pattern, while simulation~g has a unimodal pattern. As mentioned in Figure~\ref{fig:all_scatter}, simulation~g has an age gap but there is not enough star formation at $\FeH\sim0$ before the gap to result in a strong bimodality. Otherwise, while a global quiescent period is sufficient for generating an age gap, most simulations do not have a global quiescent period.

\section{Cause of Suppressed Star Formation}\label{app:cause_qui}
\begin{figure}
  \centering
  \includegraphics[width=242.26653pt]{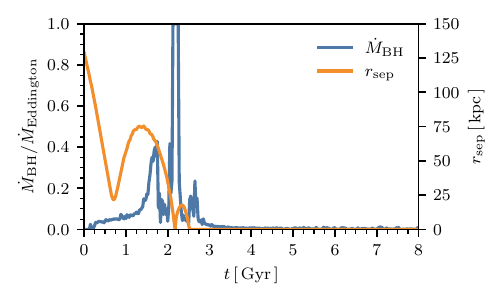}
  \caption{Evolution of black hole accretion rate and orbital separation over time in the bimodal simulation. The blue line shows the black hole accretion rate as a fraction of the Eddington rate, while the orange line shows the orbital separation between the satellite and central galaxies. The accretion rate peaks during coalescence at $\sim2\,\Gyr$, suggesting a strong connection between the merger and AGN activity.}
  \label{fig:MdotBH_rsep}
\end{figure}

In Figure~\ref{fig:MdotBH_rsep}, we demonstrate how the orbit of the bimodal simulation is closely related to the strength of BH feedback. On the $y$-axis, we show in blue the black hole accretion rate as a ratio of the maximum (Eddington) accretion rate at that time. In orange we show the orbital separation between the satellite and central galaxies. We see that the accretion rate is high early on at $\sim10\%$. At the time around coalescence at $\sim2\,\Gyr$, the accretion rate rises up to Eddington, before dropping to a much lower value $<10\%$ later on.

In the TNG model, the strength of AGN feedback is directly tied to the BH's accretion rate \citep{2017MNRAS.465.3291W}. Therefore, it is reasonable to suspect that the feedback from the AGN is responsible for removing gas from the galaxy or keeping it above the star forming density threshold.

\section{Abundance Plane of All Simulations}\label{app:allmerge}
\begin{figure*}
  \centering
  \includegraphics[width=\textwidth]{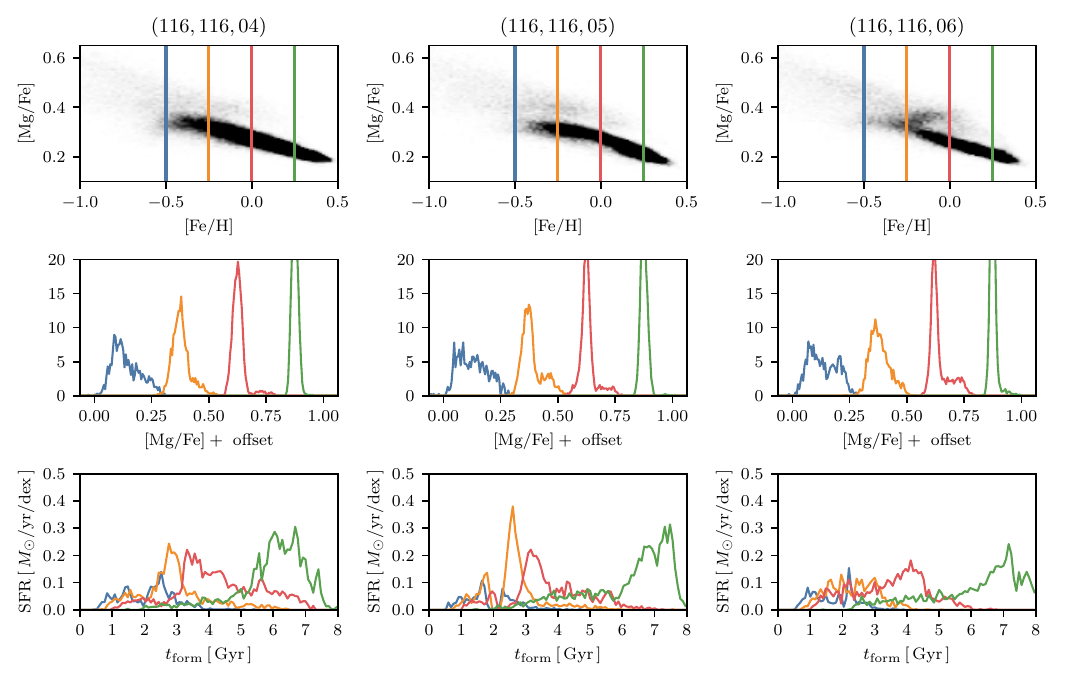}
  \caption{A summary of the abundance plane and star formation history of all simulations within the orbital grid. Each figure shows the outcome of a simulation at a fixed $R_0$ and $V_0$, varying $\eta$. The title of each column shows the $R_0$, $V_0$, and $\eta$ of that simulation, in order. The upper and middle rows replicate Figure~\ref{fig:fig1}, which show the distribution of stars in the abundance plane of \MgFe{}-\FeH{} as well as 1D histograms at a fixed \FeH{} of $-0.5$, $-0.25$, $0$, and $0.25$. The lower rows replicate Figure~\ref{fig:before_after_sfh_by_iron}, showing the star formation history at each \FeH{}.}
  \label{fig:allmerge0}
\end{figure*}

\begin{figure*}
  \centering
  \includegraphics[width=\textwidth]{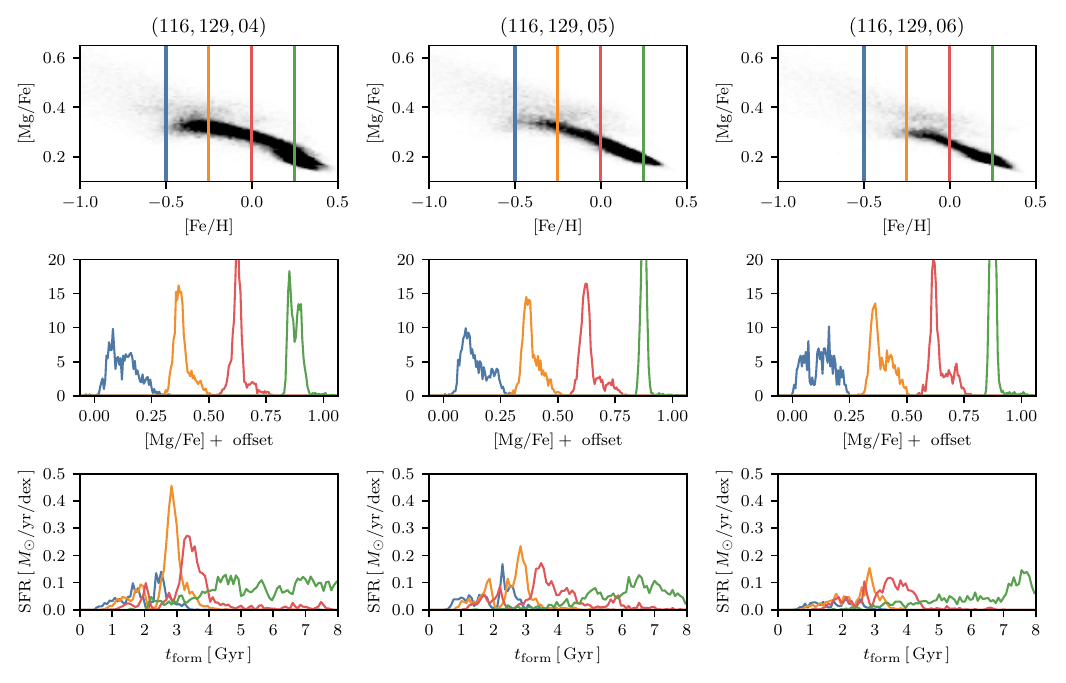}
  \caption{A continuation of Figure~\ref{fig:allmerge0}.}
  \label{fig:allmerge1}
\end{figure*}

\begin{figure*}
  \centering
  \includegraphics[width=\textwidth]{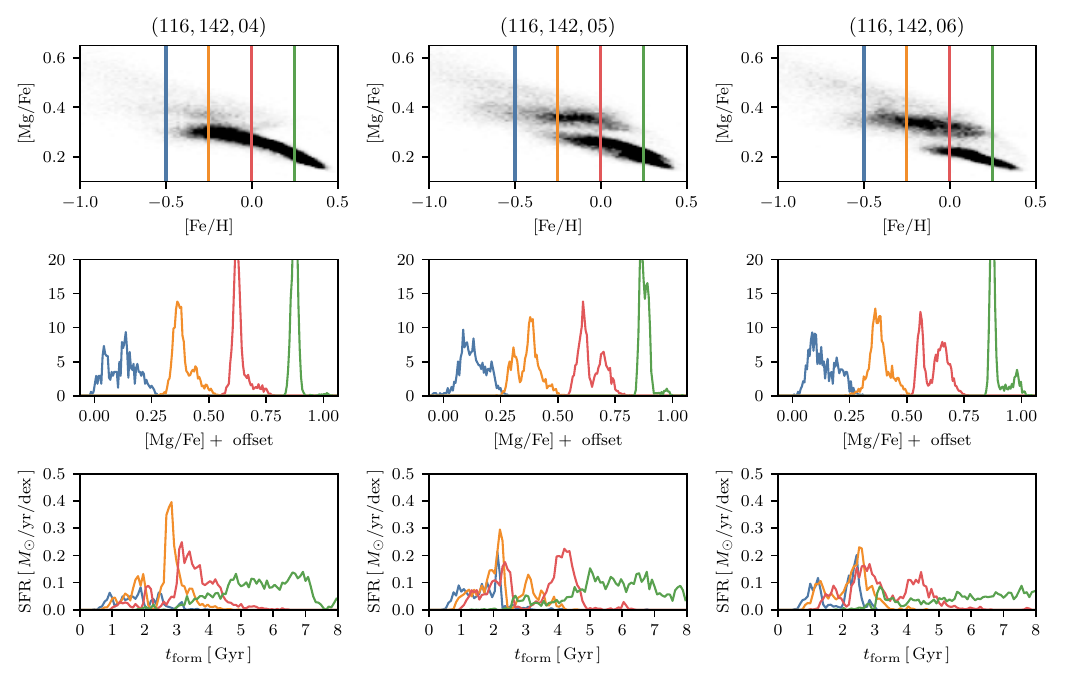}
  \caption{A continuation of Figure~\ref{fig:allmerge0}.}
  \label{fig:allmerge2}
\end{figure*}

\begin{figure*}
  \centering
  \includegraphics[width=\textwidth]{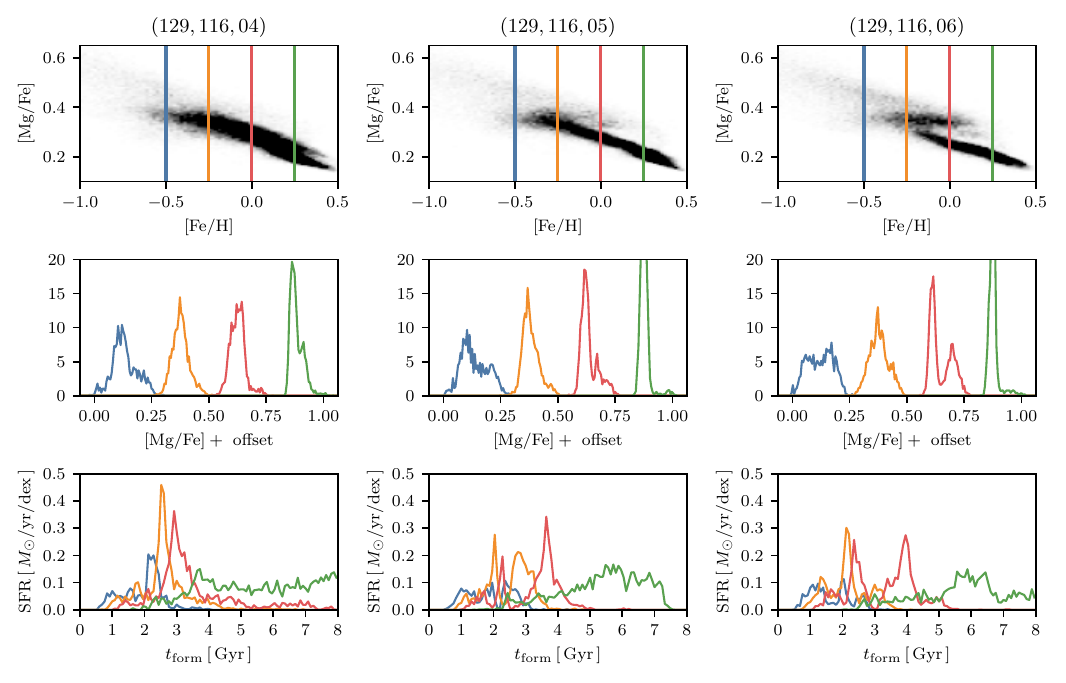}
  \caption{A continuation of Figure~\ref{fig:allmerge0}.}
  \label{fig:allmerge3}
\end{figure*}

\begin{figure*}
  \centering
  \includegraphics[width=\textwidth]{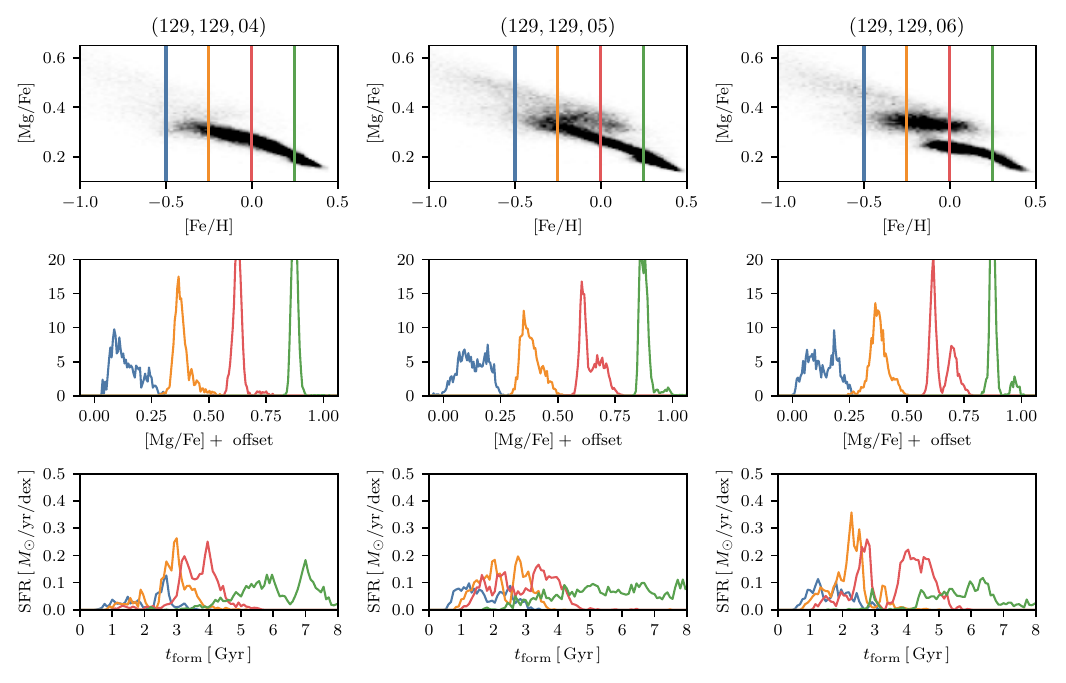}
  \caption{A continuation of Figure~\ref{fig:allmerge0}.}
  \label{fig:allmerge4}
\end{figure*}

\begin{figure*}
  \centering
  \includegraphics[width=\textwidth]{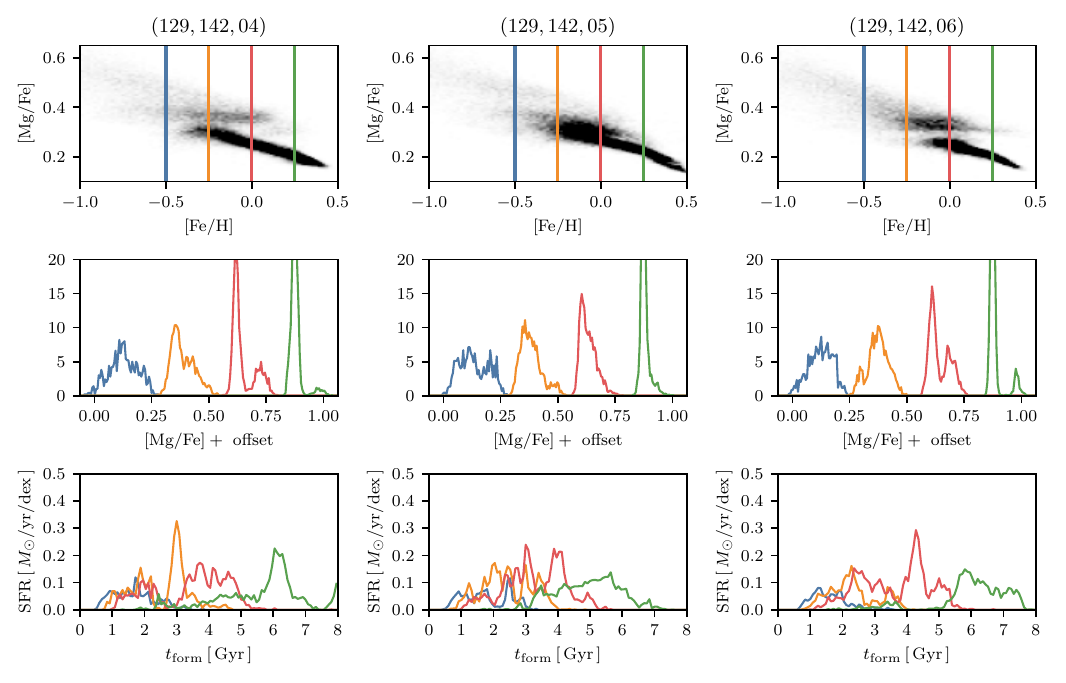}
  \caption{A continuation of Figure~\ref{fig:allmerge0}.}
  \label{fig:allmerge5}
\end{figure*}

\begin{figure*}
  \centering
  \includegraphics[width=\textwidth]{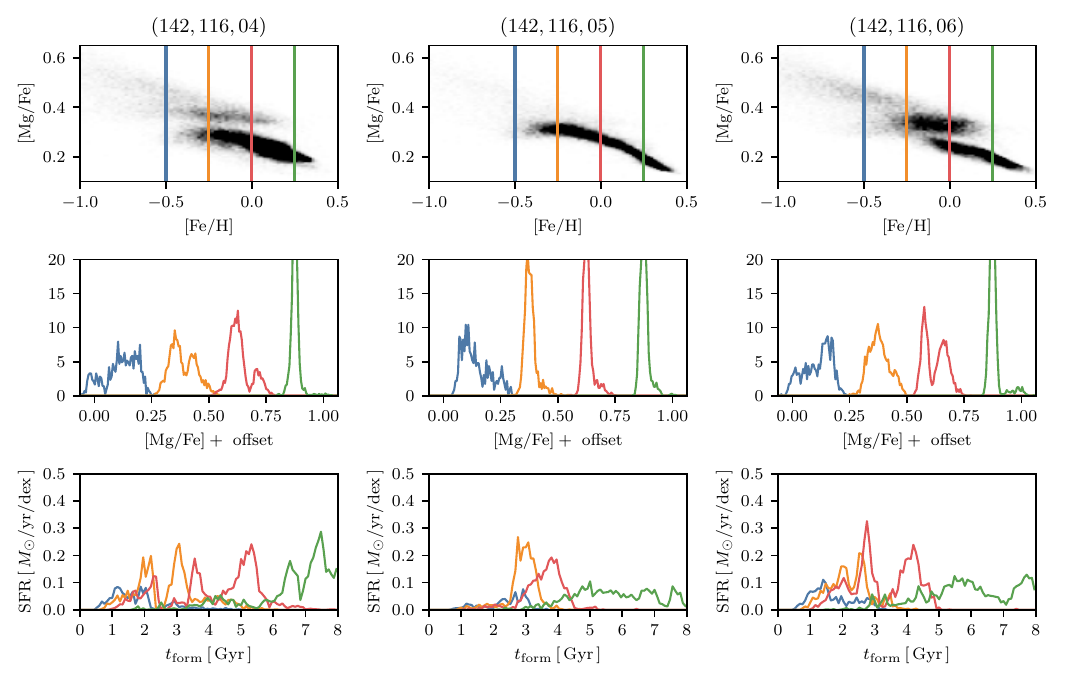}
  \caption{A continuation of Figure~\ref{fig:allmerge0}.}
  \label{fig:allmerge6}
\end{figure*}

\begin{figure*}
  \centering
  \includegraphics[width=\textwidth]{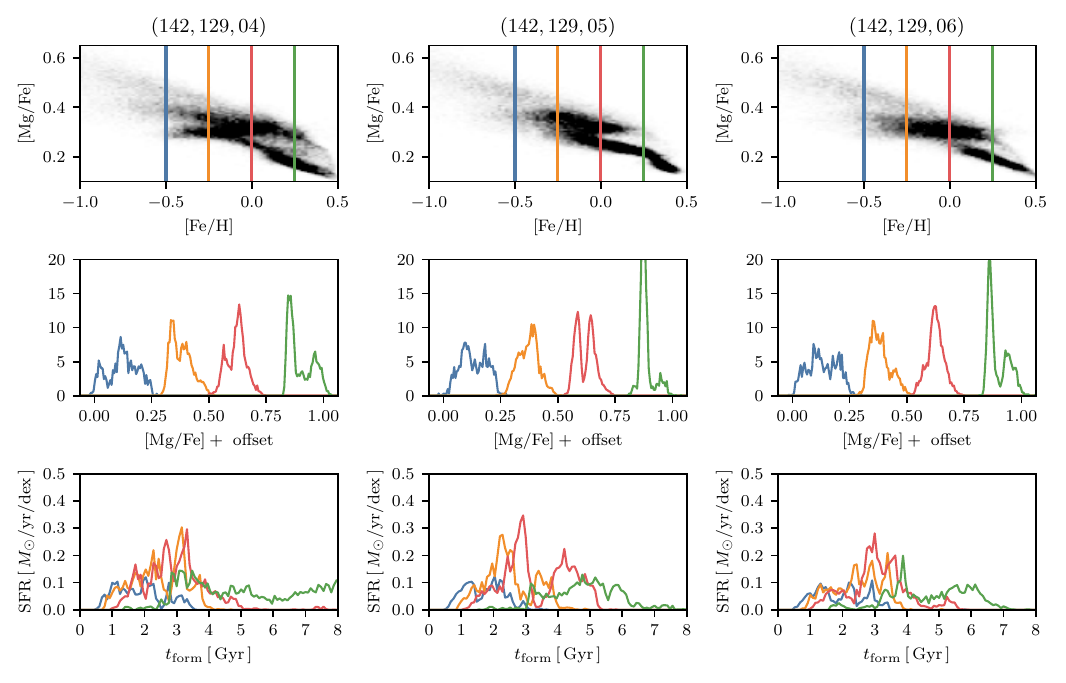}
  \caption{A continuation of Figure~\ref{fig:allmerge0}.}
  \label{fig:allmerge7}
\end{figure*}

\begin{figure*}
  \centering
  \includegraphics[width=\textwidth]{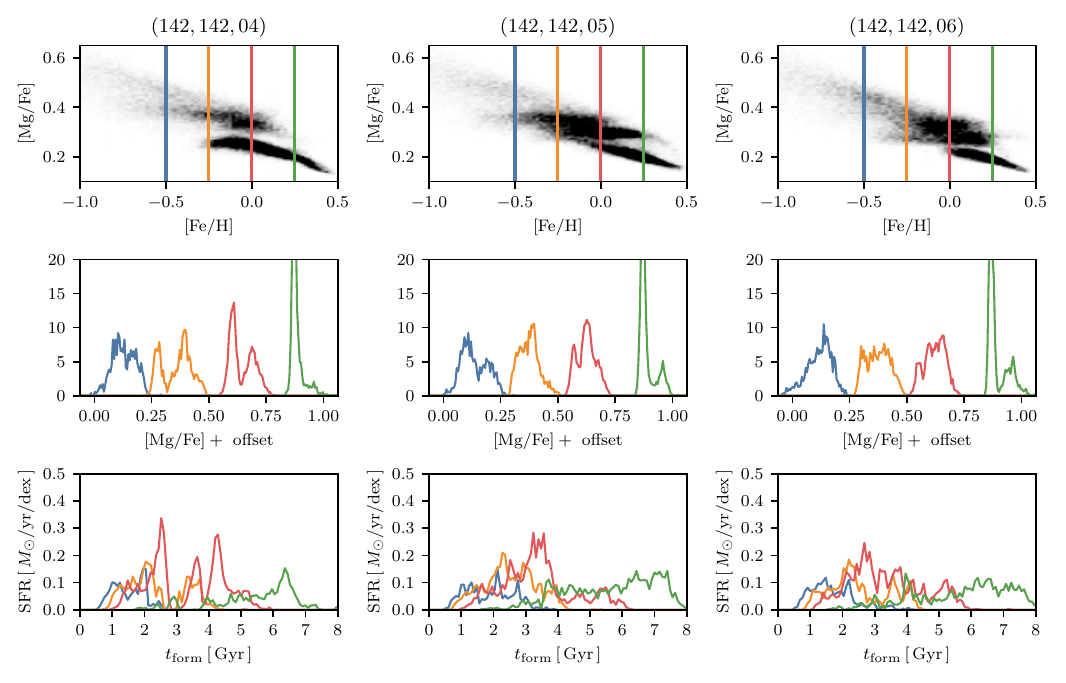}
  \caption{A continuation of Figure~\ref{fig:allmerge0}.}
  \label{fig:allmerge8}
\end{figure*}

We show summary plots of the abundance planes of all simulations in our orbital grid in Figures~\ref{fig:allmerge0} to \ref{fig:allmerge8}.

\end{document}